\documentclass[11pt,aps,prd,preprint,nofootinbib,longbibliography,superscriptaddress]{revtex4-1}

\usepackage{graphicx}
\usepackage{dcolumn}
\usepackage{bm}
\usepackage{amssymb}
\usepackage{amsmath}
\usepackage{epsfig}    
\usepackage{color}
\usepackage{slashed}
\usepackage{comment}
\usepackage{hyperref}
\begin{document} 
 
\title{Are 2HDMs with a gauged $U(1)$ symmetry alive?}

\author{Yuanchao Lou}
\email{yuanchao\_lou@nnu.edu.cn}
\affiliation{College of Physics, Sichuan University, Chengdu 610065, China}
\affiliation{Department of Physics and Institute of Theoretical Physics, Nanjing Normal University, Nanjing, 210023, China}
\author{Takaaki Nomura}
\email{nomura@scu.edu.cn}
\affiliation{College of Physics, Sichuan University, Chengdu 610065, China}
\author{Xinran Xu}
\email{xinran.xu@connect.ust.hk}
\affiliation{College of Physics, Sichuan University, Chengdu 610065, China}
\affiliation{Department of Physics, The Hong Kong University of Science and Technology, Clear Water Bay, Kowloon, Hong Kong, P.R. China}
\author{Kei Yagyu}
\email{yagyu@rs.tus.ac.jp}
\affiliation{Department of Physics, Tokyo University of Science,
1-3, Kagurazaka, Shinjuku-ku, Tokyo 162-8601, Japan}

\begin{abstract}
\noindent
We investigate the phenomenology of 2 Higgs doublet models (2HDMs) with a new $U(1)$ gauge symmetry, $U(1)_X$, by which flavor changing neutral currents are forbidden at tree level. 
As an important consequence of the spontaneous breaking of both the $U(1)_X$ and electroweak symmetries by electroweak vacuum expectation values, 
upper limits appear on masses of an additional gauge boson $Z'$ and extra Higgs bosons which are less than the TeV scale. 
In addition, the standard model (SM) like Higgs boson $h$ and a heavier Higgs boson $H$ mainly decay into a pair of $Z'$ which induces four lepton final states. These new decay modes cannot be suppressed by taking no $Z$-$Z'$ mixing and/or the Higgs alignment limit.   
We find that the minimum setup of these 2HDMs has been excluded by current data for four lepton searches at LHC. 
Such severe constraints can, however, be avoided by introducing a pair of vector-like fermions $\chi$ which are singlet under the SM symmetry but charged under $U(1)_X$, and can be a candidate of dark 
matter. Thanks to the existence of $\chi$, $Z'$ can mainly decay into $\chi\bar{\chi}$ instead of SM leptons. 
As benchmark models, 
we consider the $U(1)_H$ and $U(1)_R$ models realized by fixing specific $U(1)_X$ charges, and find regions of parameter space allowed by theoretical and current experimental constraints. We clarify that $m_H \in [160, 220]$ GeV and $\tan \beta \in [3, 4.4]$ are allowed in the $U(1)_H$ model, while $m_H \in [160, 380]$ GeV and $\tan \beta \in [1.6, 4.4]$ are allowed in the $U(1)_R$ model. 
In both the models, the $Z'$ mass is constrained to be $100~\text{GeV} \lesssim m_{Z'} \lesssim 110$ GeV. 
Such a quite limited parameter space can further be explored at future collider experiments, e.g., High-Luminosity LHC and lepton colliders. 

\end{abstract}

\maketitle
%\tableofcontents
\newpage

\section{Introduction}

The gauge principle has been remarkably successful in describing the fundamental forces for elementary particles--namely, the electromagnetic, strong, and weak interactions. 
Gravity is also expected to be governed by a gauge symmetry~\cite{Utiyama:1956sy}, although its experimental confirmation of this idea has not been confirmed yet. A natural question that follows is: Are there any fifth forces in nature?

The simplest possibility for such a fifth force is the existence of a new $U(1)$ gauge symmetry, which we denote by $U(1)_X$. 
In fact, various new physics scenarios predict a $U(1)$ gauge symmetry such as grand unified theories, e.g., $E_6$ and $SO(10)$~\cite{Hewett:1988xc}. In addition, right-handed neutrinos are naturally introduced in models with $U(1)_X$ due to the gauge anomaly 
%cancellation~\cite{Pati:1973uk, Davidson:1978pm, Marshak:1979fm}. 
cancellation~\cite{Pati:1973uk,Marshak:1979fm}. 
Furthermore, we can construct models beyond the Standard Model (SM) which can explain not only the neutrino masses but also dark matter and baryon asymmetry of the Universe, e.g., \cite{Matsui:2023bwa}. 

One of the most important consequences of introducing the $U(1)_X$ symmetry is the appearance of a new gauge boson $Z'$. Because the massless $Z'$ has highly been constrained by the tests of equivalence principle~\cite{Heeck:2014zfa}, we consider the massive $Z'$ case. The mass of $Z'$ is most economically generated by a Vacuum Expectation Value (VEV) of a scalar field which is neutral under the SM gauge symmetry but charged under $U(1)_X$. 
This scenario is compatible with the seesaw mechanism, and thus a number of phenomenological studies have been performed so far, see Refs.~\cite{Fabbrichesi:2020wbt, Langacker:2008yv} as review papers.   
However, there is, {\it a priori}, no reason to choose the Higgs sector to be one SM Higgs doublet and a singlet fields, but we can consider the other possibilities as long as the structure of the spontaneous breaking $SU(2)_L \times U(1)_Y \times U(1)_X \to U(1)_{\rm em}$ is maintained.  

In this paper, we investigate models with $U(1)_X$ which is spontaneously broken by VEVs of two Higgs doublet fields instead of a singlet scalar field. Because the current measurements of the Higgs properties at the LHC~\cite{ATLAS:2022vkf, CMS:2022dwd} indicate the existence of at least one Higgs doublet field, the existence of the second Higgs doublet seems to be plausible. Although 2 Higgs doublet models (2HDMs) suffer from the appearance of the Flavor Changing Neutral Currents (FCNCs) at tree level due to flavor violating couplings of neutral Higgs bosons, our scenario naturally suppress such dangerous FCNCs due to the $U(1)_X$ symmetry~\cite{Ko:2013zsa, Ko:2012hd, Camargo:2018uzw,Jung:2023ckm,Bae:2024lov}. 
One of the remarkable differences from the models with a singlet scalar boson is the appearance of upper limits on the masses of $Z'$ and extra Higgs bosons, typically less than the TeV scale due to the spontaneous breaking of $U(1)_X$ and the electroweak symmetry by the electroweak VEVs. Another important difference can be seen in the Higgs boson decays. 
 We find that the couplings of the 125 GeV Higgs boson ($h$) and a heavier Higgs boson $H$ with a $Z'$ pair does not vanish at the Higgs alignment limit and/or no $Z$-$Z'$ mixing limit. 
As a consequence, the branching ratio of $h$ can significantly be modified from the SM prediction, and the cross section of four lepton final states is enhanced by the decays of $h/H \to Z'Z^{\prime (*)} \to 4\ell$. By combining them, we clarify that the minimal setup of 2HDMs with $U(1)_X$ has been excluded by the current LHC data. 

We thus introduce a pair of vector-like fermions $\chi$ which are singlet under the SM symmetry but charged under $U(1)_X$, and they can be a candidate of dark matter as their stability is guaranteed by $U(1)_X$. The main decay mode of $Z'$ can then turn out to be the $\chi\bar{\chi}$ final state instead of SM leptons, so that we can avoid the constraints from 4 lepton events via the Higgs decays. 
We find the parameter space allowed by the theoretical bounds and current experimental data in the two benchmark models, i.e., the $U(1)_H$ and $U(1)_R$ models. 

This paper is organized as follows. 
In Sec.~\ref{sec:model}, we give the Lagrangians for the gauge sector and the Higgs sector, and provide the relevant interaction terms. 
In Sec.~\ref{sec:decay}, we discuss the new decay modes of the SM-like Higgs boson and the extra Higgs bosons. 
Sec.~\ref{sec:const} is devoted for the discussion of theoretical bounds and experimental constraints. 
We then show the parameter space allowed by all the constraints in Sec.~\ref{sec:combined}. 
Conclusion is given in Sec.~\ref{sec:conclusions}.

\section{Model \label{sec:model}} 

\begin{table}[t]
  \begin{center}
    \begin{tabular}{|c|c|c|c|c|c|c|c||c|c|}\hline
      &
      $~~Q_L^i~~$ & 
      $~~u_R^i~~$ & 
      $~~d_R^i~~$ & 
      $~~L_L^i~~$ & 
      $~~e_R^i~~$ & 
      $~~\nu_{R}^i~~$ &
      $~~\chi_{L,R}^{}~~$ &
      $\Phi_1$ & $\Phi_2$   
      \\ \hline
      $~~SU(2)_L~~$ & $\bf{2}$ & $\bf{1}$ & $\bf{1}$ & 
      $\bf{2}$ & $\bf{1}$ & $\bf{1}$ &$\bf{2}$ &  $\bf{2}$  & $\bf{1}$  \\ \hline
      $~~U(1)_Y~~$ & $\frac{1}{6}$ & $\frac{2}{3}$ & $-\frac{1}{3}$ &  $-\frac{1}{2}$ & $-1$ & $0$ & $0$ & $\frac{1}{2}$ & $\frac{1}{2}$  \\ \hline
      $~~U(1)_X~~$ & $\frac{1}{3}(x_2-x_R^{})$ & $\frac{1}{3}(4x_2-x_R^{})$ & $\frac{1}{3}(-2x_2-x_R^{})$ & $-x_2+x_R^{}$ & $-2x_2+x_R^{}$ & $x_R^{}$& $x_\chi$ & $x_1$& $x_2$ \\ \hline
    \end{tabular}   
  \end{center}
  \caption{Particle content and its charge assignment under the gauge symmetry $SU(2)_L\times U(1)_Y \times U(1)_X$, where charges for fermions are taken to be flavor universal. }
  \label{tab:charge}
\end{table}

We consider models with a new $U(1)_X$ gauge symmetry which is spontaneously broken by VEVs of isospin Higgs doublets $\Phi_1$ and $\Phi_2$. In order to satisfy the anomaly cancellation, we introduce three right-handed neutrinos $\nu_R^i~(i=1,\dots,3)$. 
As we discuss below, this minimal model is now excluded by constraints from Higgs boson decays, so that we introduce a pair of SM singlet vector-like fermions $\chi$ which play an important role to avoid the constraints, and can also be a dark matter candidate.  
The particle content is shown in Table~\ref{tab:charge}, in which we show the $U(1)_X$ charges, defined as $\tilde{X}_\Psi$, for a particle $\Psi$ in addition to the weak isospin $SU(2)_L$ and the hypercharge $U(1)_Y$. 
There are four free parameters for $\tilde{X}_\Psi$, which can be chosen to be $\{\tilde{X}_{\Phi_1},\tilde{X}_{\Phi_2},\tilde{X}_{\nu_R}, \tilde{X}_\chi \} = \{x_1,x_2,x_R^{}, x_\chi\}$. 
We take $x_1\neq x_2$ such that the two Higgs doublets have different charges with each other, and thus the FCNCs mediated by Higgs bosons are forbidden.  

\subsection{Gauge-kinetic term}

The most general gauge kinetic term is given by 
\begin{align}
\mathcal{L}_{\text{gauge}} &= -\frac{1}{4} W^a_{\mu \nu} W^{a \mu \nu} 
- \frac{1}{4}(\tilde{B}_{\mu\nu},\tilde{X}_{\mu\nu})
\left(
\begin{array}{cc}
1 & \sin\epsilon\\
\sin\epsilon & 1
\end{array}
\right)
\left(
\begin{array}{c}
\tilde{B}^{\mu\nu}\\
\tilde{X}^{\mu\nu}
\end{array}\right),  \label{kin} 
\end{align} 
where $W_{\mu \nu}^a$, $\tilde{B}_{\mu\nu}$ and $\tilde{X}_{\mu\nu}$ are respectively the field strength tensors of $SU(2)_L$, $U(1)_Y$ and $U(1)_X$ gauge fields, respectively, and $\sin\epsilon$ denotes the kinetic mixing parameter. 

We can move to the new basis without the kinetic mixing through the following non-unitary transformation~\cite{Holdom:1985ag}, 
\begin{align}
\left(
\begin{array}{c}
\tilde{B}_\mu\\
\tilde{X}_\mu
\end{array}\right)=
\left(
\begin{array}{cc}
1 & -\tan\epsilon\\
0 & \sec\epsilon
\end{array}\right)
\left(
\begin{array}{c}
B_\mu\\
X_\mu
\end{array}\right). \label{eq:non-uni}
\end{align}
In this basis, the kinetic terms take the diagonal form, and the covariant derivative  is then expressed as 
\begin{align}
D_\mu\Psi = (\partial_\mu -igT_\Psi^aW_\mu^a -ig'Y_\Psi B_\mu -ig_X^{}X_\Psi X_\mu )\Psi, 
\end{align}
where 
\begin{align}
X_\Psi &= \tilde{X}_\Psi - Y_\Psi\frac{g'}{g_X^{}}\tan\epsilon. \label{eq:x-charge}
\end{align}
Three neutral gauge bosons ($W_\mu^3,B_\mu,X_\mu$) are mixed after the electroweak symmetry breaking and their linear combinations would be the massless photon $A_\mu$ and massive gauge bosons $Z$ and $Z'$, see Eq.~(\ref{eq:neutral}). 

\subsection{Higgs sector}

In this subsection, we give the Lagrangian related to the Higgs fields, i.e., the Higgs potential, the Yukawa interactions and the kinetic term. 

The Higgs potential is generally expressed as 
\begin{align}
V = m_1^2 |\Phi_1|^2 + m_2^2 |\Phi_2|^2 + \frac{\lambda_1}{2}|\Phi_1|^4
+ \frac{\lambda_2}{2}|\Phi_2|^4
+ \lambda_3|\Phi_1|^2|\Phi_2|^2
+ \lambda_4|\Phi_1^\dagger \Phi_2|^2, 
\end{align}
where all the parameters are real. 
The mixing terms $\Phi_1^\dagger \Phi_2$ and $(\Phi_1^\dagger \Phi_2)^2$
are forbidden due to the $U(1)_X$ gauge symmetry, so that this potential corresponds to the special case of that given in 2HDMs with a softly-broken $Z_2$ symmetry~\cite{Glashow:1976nt}. 
We can parameterize the two Higgs doublet fields as 
\begin{align}
\Phi_i = 
\begin{pmatrix}
    \omega_i^+ \\
    \frac{h_i + v_i + iz_i}{\sqrt{2}}
\end{pmatrix}~~~(i=1,2), 
\end{align}
where $v_1$ and $v_2$ are the VEVs of the doublets which satisfy the relation $v^2 \equiv v_1^2 + v_2^2 = (\sqrt{2}G_F)^{-1}$ with $G_F$ being the Fermi constant. 
In the above expression, both $z_1$ and $z_2$ become the Nambu-Goldstone (NG) bosons which turn out to be the longitudinal components of the $Z$ and $Z'$ bosons. 
Thus, there is no physical CP-odd Higgs boson, normally denoted as $A$, in our model. 
The charged NG bosons $G^\pm$ and the physical charged-Higgs bosons $H^\pm$ can be identidied by the following orthogonal transformation:  
\begin{align}
\begin{pmatrix}
\omega_1^\pm\\
\omega_2^\pm
\end{pmatrix}
=
R(\beta)
\begin{pmatrix}
G^\pm\\
H^\pm
\end{pmatrix}, 
\end{align}
where $\tan\beta = v_2/v_1$, and 
\begin{align}
R(\theta) = 
\begin{pmatrix}
c_\theta & -s_\theta \\
s_\theta & c_\theta
\end{pmatrix}, 
\end{align}
with $c_\theta = \cos\theta$, $s_\theta = \sin\theta$ and $t_\theta = \tan\theta$. 
The remaining two CP-even components $h_1$ and $h_2$ 
can be mixed with each other as follows
\begin{align}
    \begin{pmatrix}
        h_1 \\
        h_2
    \end{pmatrix}=
   R(\alpha)
    \begin{pmatrix}
        H \\
        h
    \end{pmatrix},   
\end{align}
where we identify the $h$ state as the discovered Higgs boson with the mass of 125 GeV, while $H$ is the additional Higgs boson. The mixing angle $\alpha$ is determined by Eq.~(\ref{eq:alpha}) given below. 

By solving the tadpole conditions, i.e., 
\begin{align}
T_1 &= \frac{\partial V}{\partial h_1}\Bigg|_0 
\leftrightarrow
m_1^2 = -\frac{v^2}{2}  \left[ \lambda
_1c^2_\beta +(\lambda_3+\lambda_4)s^2_\beta  \right],\\
T_2 &\equiv \frac{\partial V}{\partial h_2}\Bigg|_0 
\leftrightarrow 
   m_2^2 = -\frac{v^2}{2}  \left[
   \lambda_
2 s^2_\beta +
   (\lambda_3+\lambda_4)c^2_\beta \right],
\end{align}
the parameters $m_1^2$ and $m_2^2$ are determined by the VEV and $\lambda_i$ parameters. 
The mass of $H^\pm$ is then given by 
\begin{align}
m_{H^\pm}^2 = -\frac{v^2}{2}\lambda_4. 
\end{align}
The mass matrix for the neutral Higgs bosons is given in the basis of ($h_1,h_2$) as 
\begin{align}
    M_{\rm neut}^2 = v^2
    \begin{pmatrix}
        \lambda_1 c^2_\beta & (\lambda_3 + \lambda_4)c_\beta s_\beta \\
        (\lambda_3 + \lambda_4)c_\beta s_\beta & \lambda_2 s^2_\beta
    \end{pmatrix}, 
\end{align}
and then the masses of $h$ and $H$ and the mixing angle $\alpha$ are expressed as 
\begin{align}
    m_h^2 &= v^2\left[\lambda_2 s^2_\beta c^2_\alpha + \lambda_1 c^2_\beta s^2_\alpha -\frac{1}{2}(\lambda_3 + \lambda_4) s_{2\beta} s_{2\alpha}\right],\\
        m_H^2 &= v^2\left[\lambda_2 s^2_\beta c^2_\alpha + \lambda_1 c^2_\beta s^2_\alpha +\frac{1}{2}(\lambda_3 + \lambda_4) s_{2\beta} s_{2\alpha}\right], \\
        t_{2\alpha} &= \frac{(\lambda_3+\lambda_4)t_\beta}{\lambda_1 - \lambda_2t^2_\beta}. \label{eq:alpha}
\end{align}
All these masses are proportional to the VEV, so that they cannot be taken to be arbitrary large values. In practice, their masses are typically constrained to be smaller than about 700-800 GeV from the perturbative unitarity bound, see Sec.~\ref{sec:const} .

Now, the parameters in the potential can be rewritten in terms of the following six physical parameters, 
\begin{align}
\{m_h,~m_H,~m_{H^\pm},~t_{\alpha},~t_\beta,~v\}, \label{eq:parameters}
\end{align}
among which $m_h\simeq 125$ GeV and $v\simeq 246$ GeV are determined by the experimental values, so that 
there are four free parameters in the potential. 
The original $\lambda_i$ ($i=1,\dots, 4$) parameters can be expressed by the parameters shown in Eq.~(\ref{eq:parameters}) as follows:     
\begin{align}
    v^2\lambda_1 &= (m_h^2 +  m_H^2t^2_\beta)s^2_{\beta-\alpha} + (m_h^2t^2_\beta + m_H^2)c^2_{\beta-\alpha} - (m_h^2 - m_H^2)s_{2(\beta-\alpha)}t_\beta, \label{eq:lam1}\\
    v^2\lambda_2 &= (m_h^2 +  m_H^2/t^2_\beta)s^2_{\beta-\alpha} + (m_h^2/t^2_\beta + m_H^2)c^2_{\beta-\alpha} + (m_h^2 - m_H^2)s_{2(\beta-\alpha)}/t_\beta, \\
    v^2\lambda_3 &= 2m_{H^\pm}^2 +  (m_H^2-m_h^2)(s_{\beta-\alpha} + c_{\beta-\alpha}/t_\beta)(s_{\beta-\alpha} - c_{\beta-\alpha}t_\beta), \\
    v^2\lambda_4 &= -2m_{H^\pm}^2. \label{eq:lam4}
\end{align}

Thanks to the $U(1)_X$ symmetry, we can naturally realize the Yukawa interactions without the tree level FCNCs, namely, only $\Phi_2$ couples to the fermions as follows
\begin{align}
{\cal L}_Y = -Y_u\bar{Q}_L \tilde{\Phi}_2 u_R^{}-Y_d\bar{Q}_L \Phi_2 d_R^{}-Y_e\bar{L}_L \Phi_2 e_R^{} -Y_\nu\bar{L}_L \Phi_2 \nu_R^{} +\text{h.c.}
\end{align}
This structure is the same as that given by the Type-I 2HDM. We note that the other Yukawa types such as Type-II, Type-X and Type-Y~\cite{Barger:1989fj, Grossman:1994jb, Aoki:2009ha} are not allowed in our framework, because they require additional chiral fermions which are charged under $U(1)_X$ in order to cancel the gauge anomaly~\cite{Ko:2012hd, Ko:2013zsa}. 
We also note that neutrinos obtain the Dirac type masses via the Yukawa coupling $Y_\nu$.

The kinetic terms for $\Phi_{1,2}$ are given by 
\begin{align}
{\cal L}_{\rm kin} & = 
|D_\mu \Phi_1|^2 + |D_\mu \Phi_2|^2.
\label{eq:kin}
\end{align}
We note that we cannot move to the Higgs basis~\cite{Davidson:2005cw} because $\Phi_1$ and $\Phi_2$ have different $U(1)_X$ charges with each other, and thus 
the kinetic term is not invariant under the $U(2)$ transformation of the two doublets. 
From the kinetic term, we obtain the mass matrix for three neutral gauge bosons as follows: 
\begin{align}
{\cal L}_{\rm mass} &= \frac{v^2}{8}
(W_3^\mu,B^\mu,X^\mu)
\begin{pmatrix}
g^2 & -gg' & -2gg_X^{}(c_\beta^2 X_{\Phi_1}+s_\beta^2 X_{\Phi_2})\\
& g^{\prime 2} & 2g'g_X^{}(c_\beta^2 X_{\Phi_1}+s_\beta^2 X_{\Phi_2})\\
&& 4g_X^2(c_\beta^2 X_{\Phi_1}^2 +s_\beta^2 X_{\Phi_2}^2)
\end{pmatrix}
\begin{pmatrix}
W_{3\mu}\\
B_\mu\\
X_\mu
\end{pmatrix}\notag\\
& =\frac{v^2}{8}
(A^\mu,Z^\mu,Z^{\prime \mu})
\begin{pmatrix}
0 & 0 & 0 \\
0 & m_Z^2 & 0 \\
0 & 0 & m_{Z'}^2
\end{pmatrix}
\begin{pmatrix}
A_\mu\\
Z_\mu\\
Z'_\mu
\end{pmatrix}, \label{eq:neutral}
\end{align}
where $A_\mu$, $Z_\mu$ and $Z_\mu'$ are respectively the massless photon, the $Z$ boson and the new $Z'$ boson. The relation between the original states and the mass eigenstates are expressed as 
\begin{align}
\begin{pmatrix}
W_3^\mu\\
B_\mu\\
X_\mu
\end{pmatrix}
=
\begin{pmatrix}
s_W^{} & c_W^{} & 0\\
c_W^{} & -s_W^{} & 0 \\
0 & 0 & 1
\end{pmatrix}
\begin{pmatrix}
1 & 0 & 0\\
0 & c_\zeta & -s_\zeta \\
0 & s_\zeta & c_\zeta
\end{pmatrix}
\begin{pmatrix}
A^\mu\\
Z_\mu\\
Z'_\mu
\end{pmatrix},  
\end{align}
where $c_W^{} = \cos\theta_W^{}$ and $s_W^{} = \sin\theta_W^{}$ and $\theta_W$ being the weak mixing angle.
In the above, the first rotation matrix separates the photon field from the two other massive gauge fields, while the second rotation diagonalizes the mass matrix, transforming the two massive states into their mass eigenstates.  
The squared masses of $Z$ and $Z'$ bosons and the mixing angle $\zeta$ are given by 
\begin{align}
m_{Z,Z'}^2 & = M_{11}c_\zeta^2 + M_{22}s_\zeta^2 \pm M_{12}s_{2\zeta}, \\
t_{2\zeta} & = \frac{2M_{12}}{M_{11} - M_{22}}, 
\end{align}
with 
\begin{align}
M_{11}&=\frac{g^2}{4c_W^2}v^2,~\\
M_{22}&=g_X^2v^2(c_\beta^2 X_{\Phi_1}^2 + s_\beta^2 X_{\Phi_2}^2 ),~\\
M_{12}&=-\frac{gg_X^{}}{2c_W^{}}v^2(c_\beta^2 X_{\Phi_1} + s_\beta^2 X_{\Phi_2}).
\end{align}
The mixing angle $\zeta$ must be much smaller than unity, typically $10^{-3}$ or smaller for $m_{Z'}< {\cal O}(100)$ GeV~\cite{Nomura:2024pwr}, 
so that the element $M_{12}$ should be suppressed. 
This can be realized by taking $g_X^{} \ll 1$ and/or $c_\beta^2 X_{\Phi_1} + s_\beta^2 X_{\Phi_2} \ll 1$. 
The first option does not work, because the $Z'$ mass becomes quite small as compared with $m_Z$, and it makes the decay of $h \to Z'Z'$ dominant, see Sec.~\ref{sec:decay}. We thus take the second option, while keeping $(c_\beta^2 X_{\Phi_1}^2 + s_\beta^2 X_{\Phi_2}^2 )$ to be order 1. 
For simplicity, let us impose the following condition to cancel the $Z$-$Z'$ mixing: 
\begin{align}
c_\beta^2 X_{\Phi_1} + s_\beta^2 X_{\Phi_2}=0 \leftrightarrow 
\frac{X_{\Phi_1}}{X_{\Phi_2}} = -t^2_\beta. \label{eq:nomixing}
\end{align}
Under this condition, the $Z'$ mass is given by 
\begin{align}
m_{Z'}^{} = g_X^{}t_\beta |X_{\Phi_2}|v. \label{eq:mzpSq}
\end{align}

In what follows, we consider two concrete models which satisfy the no-mixing condition given in Eq.~(\ref{eq:nomixing}), namely, the hidden $U(1)_H$ model~\cite{Holdom:1985ag,Ko:2012hd} and the $U(1)_R$ model~\cite{Jung:2009jz,Ko:2012hd,Nomura:2017tih}. 
The $U(1)_H$  model is realized by taking $\tilde{X}_\Psi \to 0$ with $\Psi \neq \Phi_1$ by which $Z'$ can interact with the SM particles only via the kinetic-mixing parameter $\epsilon$. 
Using Eqs.~(\ref{eq:x-charge}), (\ref{eq:nomixing}) and (\ref{eq:mzpSq}), 
we obtain  
\begin{align}
t_\epsilon = \frac{2x_1g_X^{}}{g'(1 + t^2_\beta)},\quad 
m_{Z'}^{} = \frac{g'}{2}t_\beta t_\epsilon v. 
%\frac{t_\beta}{1 + t^2_\beta}x_1g_X^{}v. 
\label{eq:no-mixing-U1H}
\end{align}
The $U(1)_R$ model is realized by taking $x_2^{} = x_R^{}$, where the $\epsilon$ parameter is taken to be zero for simplicity. 
In this model, we obtain $X_\Psi \to 0$ for $\Psi=Q_L$ and $L_L$, see Table~\ref{tab:charge}. The no-mixing condition and $Z'$ mass are then simply given by   
\begin{align}
\frac{x_1}{x_R} = -t^2_\beta,\quad
m_{Z'}^{} = g_X^{} x_R t_\beta v. \label{eq:no-mixing-U1R}
\end{align}
We note that the other $U(1)_X$ models, including $U(1)_{B-L}$,  can also realize the above condition as long as we have two independent charges $x_1$ and $x_2$. However, the other models are highly constrained by the neutrino scattering experiments as we will discuss below.

\begin{figure}[t]
 \begin{center}
\includegraphics[width=8cm]{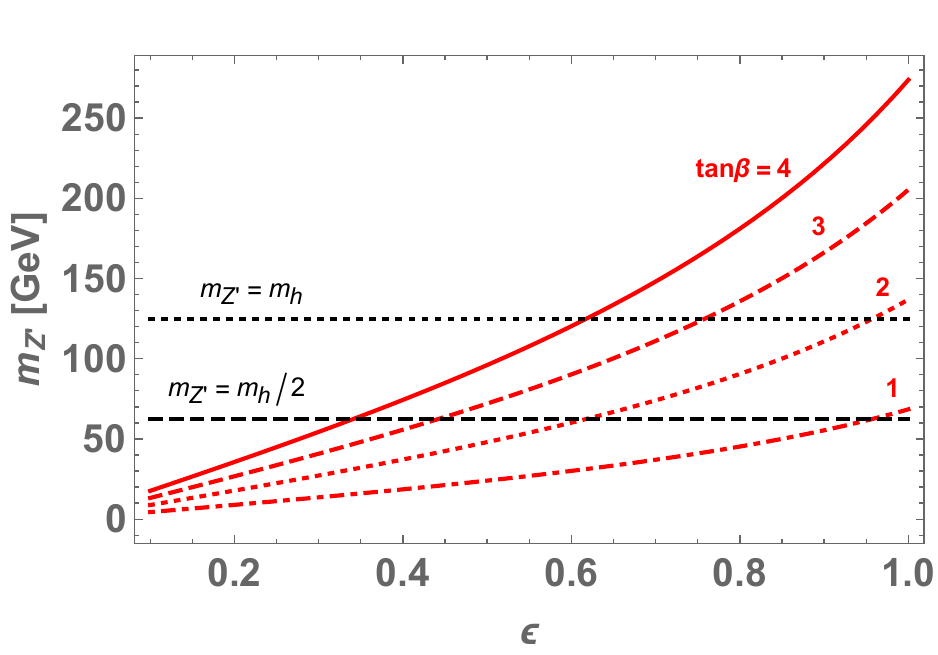}  \
\includegraphics[width=8cm]{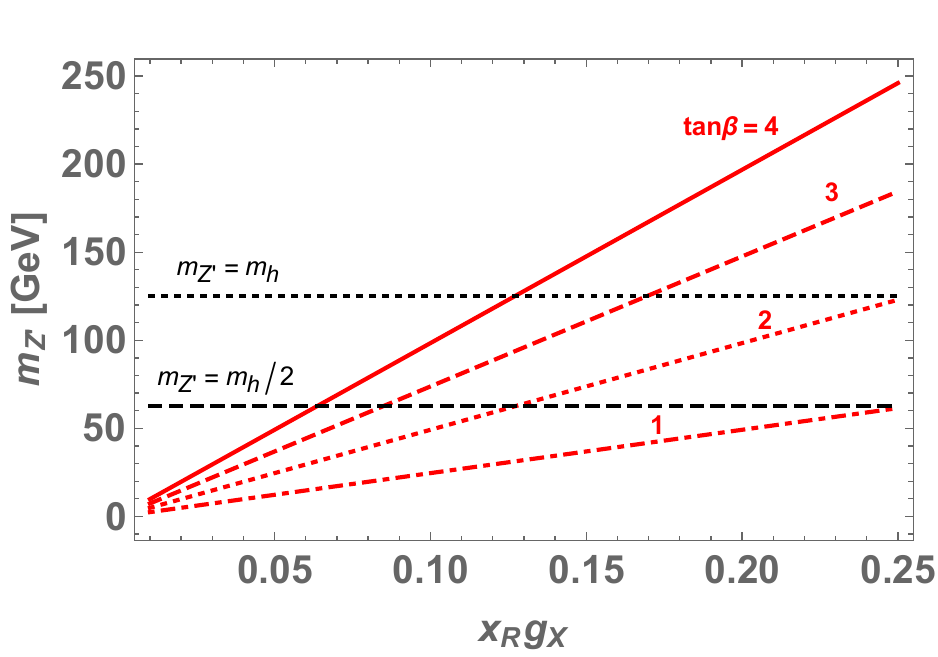}  
 \caption{Left and right plots respectively show the $Z'$ mass as a function of $\epsilon$ and $x_R g_X^{}$ for fixed values of $\tan\beta$ in the $U(1)_H$  and $U(1)_R$ models. }
\label{fig:zpmass}
\end{center}
\end{figure}

In Fig.~\ref{fig:zpmass}, we show $m_{Z'}$ as a function of the kinetic-mixing parameter $\epsilon$ in the $U(1)_H$ model (left) and $x_R^{}g_X$ in the $U(1)_H$ model. 

\subsection{Interactions under the no $Z$-$Z'$ mixing condition}

We here present the interaction terms of the Higgs bosons and $Z'$ under the no $Z$-$Z'$ mixing condition discussed above. 

The gauge-gauge-Higgs type interactions are extracted as follows: 
\begin{align}
{\cal L}_{\rm kin}& = \frac{m_Z^2}{v}(s_{\beta-\alpha}h + c_{\beta-\alpha}H)Z_\mu Z^\mu  + \frac{2m_W^2}{v}(s_{\beta-\alpha}h + c_{\beta-\alpha}H)W_\mu W^\mu \nonumber \\
&+ \frac{m_{Z'}^2}{v}\left(s_{\beta-\alpha} + c_{\beta-\alpha}\frac{1 -  t^2_\beta}{t_\beta}\right)Z_\mu^\prime Z^{\prime\mu}h  + \frac{m_{Z'}^2}{v}\left(c_{\beta-\alpha} - s_{\beta-\alpha}\frac{1 -  t^2_\beta}{t_\beta}\right)Z_\mu^\prime Z^{\prime\mu}H
\nonumber \\
&+ \frac{2m_Z^{}m_{Z'}}{v}(s_{\beta-\alpha}H-c_{\beta-\alpha} h )Z_\mu Z^{\prime\mu} \notag\\
& + \frac{2m_Wm_{Z'}}{v}H^+ W_\mu^- Z^{\prime \mu} + \text{h.c.} 
%& \equiv \sum_{\phi = \{h, H, H^\pm\}} \sum_{V,V' = \{W^\pm,Z,Z'\}} g_{\phi VV'} \phi V_\mu V'^\mu. 
\label{eq:int-kin}
\end{align}
It is worth to mention here that the $hZ'Z'$ and $HZ'Z'$ couplings do {\it not} vanish in the Higgs alignment limit, i.e., $c_{\beta-\alpha} \to 0$, where all the $h$ couplings with the SM particles become the same as those of the SM values at tree level. 
This property is notable difference from the models with $U(1)_X$ which is broken by the VEV of an isospin-singlet scalar boson. In the latter scenario, the $hZ'Z'$ coupling is suppressed by the mixing angle with the singlet-like Higgs boson, see e.g.~\cite{Nomura:2024pwr}.

\begin{table}[t]
  \begin{center}
    \begin{tabular}{|c|c| c c c c c c|c c c c c c|}\hline
      & $g_{ffZ'}$ & $v_{e}$ & $v_{\nu_L}$ & $v_{d}$ & $v_{u}$ & $v_{\nu_R}$ & $v_{\chi}$ & $a_{e}$ & $a_{\nu_L}$ & $a_{d}$ & $a_{u}$ & $a_{\nu_R}$ & $a_{\chi}$ \\ \hline  
      $U(1)_H$ & $\frac{1}{2} g' \tan \epsilon$ & $\frac32$ & $\frac12$ & $\frac16$ & $-\frac{5}{6}$ & $0$ & $R_1$ & $-\frac12$ & $\frac12$ & $-\frac12$ & $\frac12$ & $0$ & $0$ \\ \hline
      $U(1)_R$ & $\frac12 g_X x_R$ & $-1$ & $0$ & $-1$ & $1$ & $1$ & $R_2$ & $1$ & $0$ & $1$ & $-1$ & $-1$ & 0 \\ \hline  
    \end{tabular}   
  \end{center}
  \caption{Coupling factors of the $Z'f\bar{f}$ interactions in the $U(1)_H$ and $U(1)_R$ models under the no $Z$-$Z'$ mixing condition. Here, $R_1 \equiv (1+\tan^2\beta) x_\chi/(x_1 \cos \epsilon)$ and $R_2  \equiv 2x_\chi/x_R$.}
  \label{tab:ffZp}
\end{table}

\begin{figure}[tb]
 \begin{center}
\includegraphics[width=8cm]{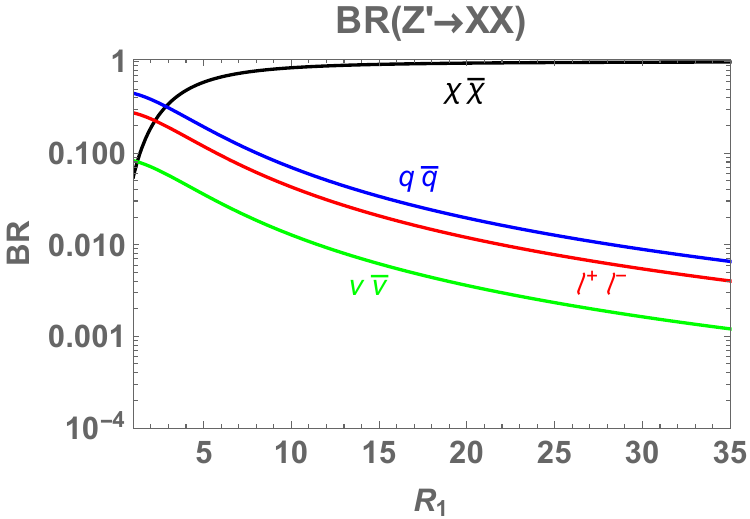}  \
\includegraphics[width=8cm]{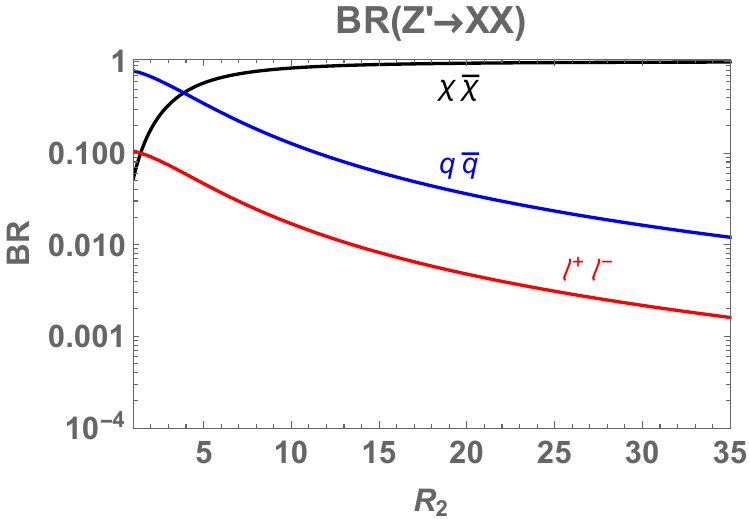}  
 \caption{Left and right plots respectively show the BRs of $Z'$ as a function of $R_1$ and $R_2$ in the $U(1)_H$ and $U(1)_R$ models. For $q \bar{q}$ ($q\neq t$), $\ell^+ \ell^-$ and $\nu \bar{\nu}$ final states, we summed all the flavors. }
\label{fig:BRzp}
\end{center}
\end{figure}

The $Z'$ couplings with a fermion pair are extracted as follows: 
\begin{equation}
\mathcal{L}_{Z'ff} = g_{ffZ'} \overline{f} \gamma^\mu (v_f - a_f \gamma_5 ) f Z'_\mu, \label{eq:zpchichi}
\end{equation}
where $g_{ffZ'}$, $v_{f}$ and $a_{f}$ are summarized in Table~\ref{tab:ffZp} for the $U(1)_H$ and $U(1)_R$ cases. 
We then obtain decay width for $Z' \to f \bar{f}$ process as 
\begin{equation}
\Gamma_{Z' \to f \bar{f}} \simeq g_{ffZ'}^2 \frac{v_f^2 + a_f^2}{12 \pi} m_{Z'},
\end{equation}
where we ignore the fermion mass.
In Fig.~\ref{fig:BRzp}, we show the decay branching ratios (BRs) of $Z'$ as a function of $R_1$ ($R_2$) in the $U(1)_H$ $(U(1)_R)$ model for $m_f \ll m_{Z'} < m_t~(f\neq t)$ where we sum up all the flavor except top quark for $q \bar{q}$, $\ell^+ \ell^-$ and $\nu \bar{\nu}$ final states.

\section{Decay of the Higgs bosons \label{sec:decay}}

In this section, we discuss the decays of $h$ and the additional Higgs bosons $H$ and $H^\pm$. In particular, we focus on the decay modes including $Z'$ in the final states. 

\subsection{$h$ decays}

In our model, $h$ can decay into new modes such as $Z'Z'$ and $Z'Z$. 
When these modes are kinematically allowed, the decay rates are given by
\begin{align}
\Gamma(h \to Z^\prime Z^\prime) 
& =\frac{m_h^3}{32 \pi v^2} 
\left(s_{\beta-\alpha} + c_{\beta-\alpha}\frac{1 -  t^2_\beta}{t_\beta} \right)^2
\left[12x_{Z'}^2+\lambda(x_{Z'},x_{Z'})\right] \lambda^{1/2}(x_{Z'},x_{Z'}), \\
\Gamma(h \to Z Z^\prime) & = \frac{m_h^3 }{{16\pi v^2}} c^2_{\beta - \alpha}\left[
12x_{Z}x_{Z'}+\lambda(x_{Z},x_{Z'})
\right]\lambda^{1/2}(x_{Z^\prime}^{}, x_Z^{}) , 
\end{align}
where $\lambda(x, y) \equiv (1-x-y)^2-4xy$ and $x_V \equiv m_V^2/m_h^2$.
We see that the decay rate of $h \to ZZ'$ vanishes in the alignment limit $c_{\beta - \alpha} \to 0$, while that of $h\to Z'Z'$ is not suppressed in this limit. For $m_{Z'}/m_h \ll 1$, the decay rate is approximately given in the alignment limit  
\begin{align}
\Gamma(h \to Z^\prime Z^\prime) 
& = \frac{m_{h}^3}{32\pi v^2 }[1 - 6x_{Z'}+ {\cal O}(x_{Z'}^2)]  \simeq 
0.3\left[1 - 0.04\left(\frac{m_{Z'}}{10~\text{GeV}}\right)^2\right]~\text{GeV}.
\end{align}
This is quite huge as compared with the total width of the SM Higgs boson, i.e., $\Gamma_h^{\rm SM} = 4.1$ MeV~\cite{LHCHiggsCrossSectionWorkingGroup:2016ypw}.
Therefore, the $h \to Z'Z'$ mode easily dominates the branching ratio of the Higgs boson, and the model is excluded by the current LHC data~\cite{ATLAS:2022vkf, CMS:2022dwd}.

We thus consider the case with $m_{Z'}>m_h/2$, where the decay of $h \to Z'Z'$ is not kinematically allowed, but $h \to Z'Z^{\prime *} \to  Z'f\bar{f}$ is possible with $f$ being a SM fermion if $m_{Z'}< m_h$. 
The decay width of such a three body decay is given by 
\begin{align}
 \Gamma(h \to Z' f \bar f) &= \frac{m_{Z'}^4 g_{ffZ'}^2 (v^2_{f}+a^2_{f})}{32 \pi^3 m_h v^2}N_f^c
 \left(s_{\beta-\alpha} + c_{\beta-\alpha}\frac{1-t_\beta^2}{t_\beta}\right)^2
 H(x_{Z'}^{},x_{Z'}^{}), 
\end{align}
where $N_f^c=3~(1)$ for $f$ being quarks (leptons) and $g_{ffZ'}$, $v_{f}$ and $a_{f}$ are given in Table~\ref{tab:ffZp}.  
The phase space function $H$ is given by 
\begin{align}
H(x, y^*) &= 
 \frac{ \arctan \left( \dfrac{1 - x - y^*}{\sqrt{-\lambda(x, y^*)}} \right)
+ \arctan \left( \dfrac{1 - x + y^*}{\sqrt{-\lambda(x, y^*)}} \right)}{
2x \sqrt{-\lambda(x, y^*)}}  \notag\\
& \times
\left[ (1 - y^*)^3 - 3x^3 + (9y^* + 7)x^2 - 5(1 - y^*)^2 x \right] 
 \notag\\
&+ \frac{1}{12xy^*} \Biggl\{
(x - 1) \left[ 6y^{*2} + y^*(39x - 9) + 2(1 - x)^2 \right]  \notag\\
& \qquad \qquad \qquad- 3y^* \left[ y^{*2} + 2y^*(3x - 1) - x(3x + 4) + 1 \right] \ln x
\Biggr\}, \label{eq:h3decay}
\end{align}
where $y^*$ in $H(x,y^*)$ corresponding to a valuable for an off-shell particle. 
For $y^* = x^*$, we obtain 
\begin{align}
H(x,x^*)&=
\frac{20x^2-8x+1}{2x\sqrt{4x-1}}\arccos\left(\frac{3x-1}{2x^{3/2}}\right)\notag\\
& +\frac{x-1}{12x^2}\left(47x^2-13x+2\right) - \frac{\ln x}{4x}(4x^2-6x+1), \label{fx-3body}
\end{align}
which is consistent with the expression given in Ref.~\cite{Keung:1984hn}. 
This function takes order unity for $x\gtrsim 1/4$ ($m_{Z'}\gtrsim m_h/2$), while it is suppressed by 
\begin{align}
H(1-\delta,1-\delta) = \frac{\delta^5}{20} + {\cal O}(\delta^6). \label{eq:off-shell}
\end{align}
From this expression, we see that the decay rate is rapidly suppressed when the mass of $Z'$ approaches to the Higgs mass, i.e., $m_{Z'} \lesssim m_h$.  

\begin{figure}[t]
 \begin{center}
\includegraphics[width=8cm]{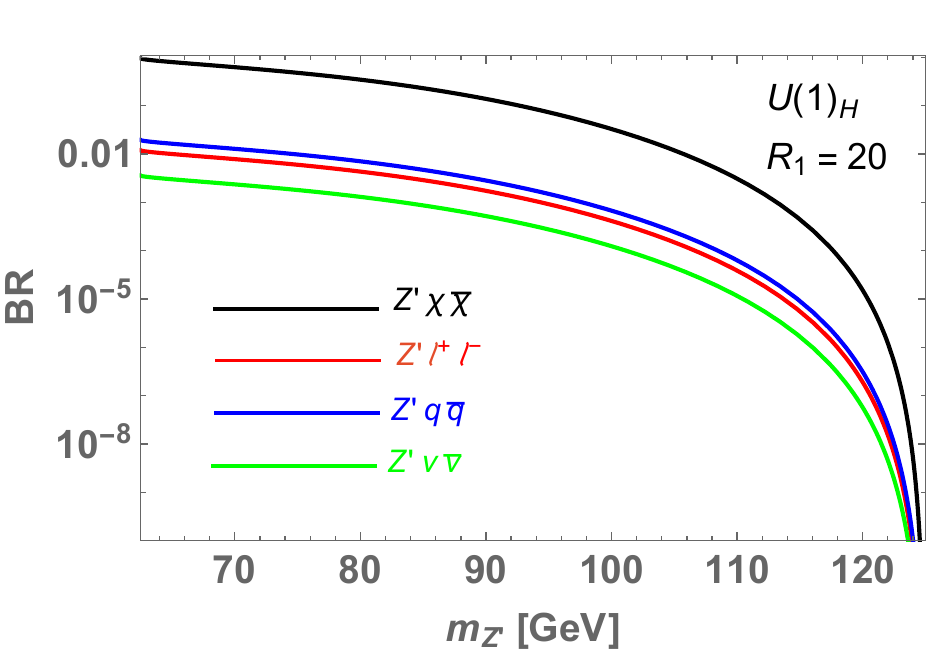}  \
\includegraphics[width=8cm]{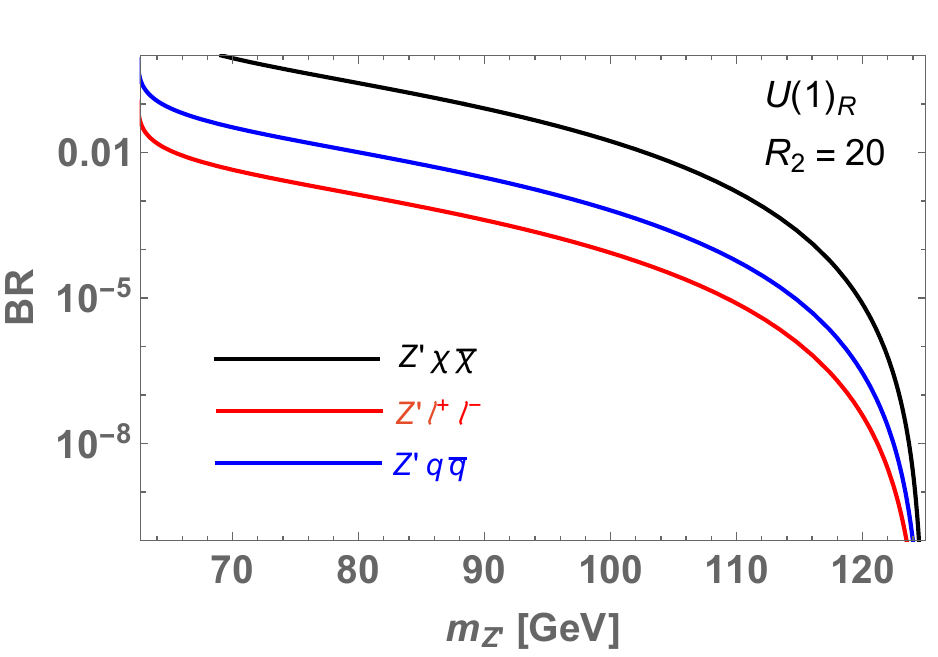}  
 \caption{Left and right plots respectively show the BRs of $h \to Z' Z'^* \to Z' f_{} \bar{f}$
 as a function of $m_{Z'}$ in the $U(1)_H$ model with $R_1=20$ and the $U(1)_R$ model with $R_2=20$ as a function of $m_{Z'}$.  All the flavors are summed up (except for the top quark) for $f \bar{f}$ final states. }
\label{fig:higgsBR}
\end{center}
\end{figure}

In Fig.~\ref{fig:higgsBR}, we show the BR of the $h \to Z' Z^{\prime *} \to Z' f\bar{f}$ modes, where the left and right plots correspond to the BR in the $U(1)_H$ model with $R_1=20$ and the $U(1)_R$ model with $R_2 = 20$ as a function of $m_{Z'}$, respectively.
%The $R_{1,2}$ parameters are defined in Table~\ref{tab:ffZp}. 
We see that $h$ mainly decays into invisible final states, i.e., $h \to Z'\chi\bar{\chi}$ in both the models due to the choice of the $R_{1,2}$ parameters, where $Z' \to \chi\bar{\chi}$ is also dominant as we can see from Fig.~\ref{fig:BRzp}. 
The BRs are larger than $10\%$ for $m_{Z'}\lesssim 90$ GeV, which is excluded by the current upper limit on the Higgs BR into unknown particles given at LHC~\cite{ATLAS:2023tkt}. On the other hand, the BRs are significantly suppressed when $m_{Z'}$ approaches to $m_h$ due to the phase space suppressions described in Eq.~(\ref{eq:off-shell}).   

\subsection{$H$ and $H^\pm$ decays}

We discuss decays of the additional Higgs bosons $H$ and $H^\pm$. 
The decay rates into a pair of SM particles are the same as those given in the Type-I 2HDM, see e.g., Ref.~\cite{Aoki:2009ha}. 
There are additional decay modes including the $Z'$ and their decay rates are given by  
\begin{align}
\Gamma(H \to Z^\prime Z^\prime) & =\frac{m^3_H}{32\pi v^2} 
\left(c_{\beta-\alpha} - s_{\beta-\alpha}\frac{1 -  t^2_\beta}{t_\beta} \right)^2
\left[12y_{Z'}^2+\lambda(y_{Z'},y_{Z'})\right] \lambda^{1/2}(y_{Z'},y_{Z'}), \\
\Gamma(H \to Z Z^\prime) & = \frac{m_H^3}{{16 \pi  v^2}}  s^2_{\beta - \alpha} \left[12y_{Z}^{}y_{Z'}^{}+ \lambda(y_{Z}^{},y_{Z'}^{})\right]\lambda^{1/2}(y_Z^{},y_{Z'}^{}),
\end{align}
where $y_X^{} \equiv m_X^2/m_H^2$. 
In the alignment limit, $s_{\beta-\alpha} \to 1$,  
all these modes do not vanish as long as these are kinematically allowed. 
For $m_{H} < 2m_{Z'}$ or $m_{H} < m_{Z} + m_{Z'}$, the massive gauge bosons can be replaced by the off-shell mode, and three body final states are realized as follows
\begin{align}
\Gamma(H \to Z' Z'^* \to Z'f \bar f) &= 
 \frac{m_{Z'}^4 g_{ffZ'}^2 (v^2_{f}+a^2_{f})}{32 \pi^3 m_H v^2}N_f^c \left(c_{\beta-\alpha} - s_{\beta-\alpha}\frac{1 -  t^2_\beta}{t_\beta} \right)^2H(y_{Z'},y_{Z'}), \\
\Gamma(H \to Z Z'^*\to Zf \bar f) &= 
 \frac{m_Z^2m_{Z'}^2 g_{ffZ'}^2 (v^2_{f}+a^2_{f})}{32 \pi^3 m_H v^2} N_f^cs_{\beta-\alpha}^2H(y_Z,y_{Z'}), \\
\Gamma(H \to Z' Z^* \to Z'f \bar f) &= \frac{m_Z^2m_{Z'}^2 g_{Z}^2 [(v_{f}^{\rm SM})^2+(a_{f}^{\rm SM})^2]}{32 \pi^3 m_H v^2}N_f^c s_{\beta-\alpha}^2H(y_{Z'},y_Z), 
 %&= \frac{g_{HZZ'}^2 g_Z^2 (v^2_{Zf}+a^2_{Zf})}{128 \pi m_H} H(y_{Z'},y_Z),
\end{align}
where $H(x,y)$ is given in Eq.~\eqref{eq:h3decay}, 
and $v_f^{\rm SM} = I_f/2 -s^2_W Q_f$ and $a_f^{\rm SM}=I_f/2$ with $I_f$ ($Q_f$) being the isospin (electric charge) of a fermion $f$.
When $m_H$ is greater than $m_{H^\pm}$, $H$ can also decay into $H^\pm$ and a $W$ boson where the $W$ boson is typically off-shell because a larger mass difference is excluded by the constraint from the $T$ parameter as it will be discussed below. The decay rate is given by 
\begin{align}
\Gamma(H \to H^\pm W^{\mp *}) &\equiv 2\sum_f\Gamma(H \to H^+ W^{-*}\to H^+ f\bar{f}')=
\frac{9g^4}{256\pi^3}s_{\beta-\alpha}^2m_H G\left(y_{H^\pm},y_W^{*}\right), 
\end{align}
where
 \begin{align}
G(x,y^*)&=\frac{1}{12y^*}\Bigg\{2\left(x-1\right)^3-9\left(x^2-1\right)y^*+6\left(x-1\right)y^{*2}\notag\\
&+6\left(1+x-y^*\right)y^*\sqrt{-\lambda(x,y^*)}
\left[\arctan\left(\frac{-1+x-y^*}{\sqrt{-\lambda(x,y^*)}}\right)+\arctan\left(\frac{-1+x+y^*}{\sqrt{-\lambda(x,y^*)}}\right)\right]\notag\\
&-3\left[1+\left(x-y^*\right)^2-2y^*\right]y^*\log x\Bigg\}. 
\end{align}

For $H^\pm$, they can decay into $W^\pm Z'$ at tree level in addition to the decays into a pair of SM fermions. The decay rate of $H^\pm \to W^\pm Z'$ is given by 
\begin{equation}
\Gamma(H^\pm \to W^\pm Z^{\prime}) =\frac{m_{H^\pm}^3}{16 \pi v^2}   \left[12z_W z_{Z^{\prime}} +  \lambda(z_W,z_{Z^\prime}) \right]\lambda^{1/2}(z_W,z_{Z^\prime}), 
\end{equation}
where $z_X^{} \equiv m_X^2/m_{H^\pm}^2$.
Similar to the $H$ decay, we can consider the decay modes with one of the gauge bosons being off-shell as 
\begin{align}
\Gamma(H^\pm \to W^\pm Z^{\prime *} \to W^\pm f\bar f) &=
 \frac{m_W^2m_{Z'}^2 g_{ffZ'}^2 (v^2_{f}+a^2_{f})}{32 \pi^3 m_{H^\pm}  v^2} N_f^cH(z_W^{},z_{Z'}^{}), \\
 \Gamma(H^\pm \to  Z^{\prime}W^{\pm *} \to Z' f\bar f') &=
 \frac{m_W^2m_{Z'}^2 g^2 }{64 \pi^3 m_{H^\pm} v^2}N_f^cH(z_{Z'}^{},z_W^{}). 
\end{align}
In addition, the decay of $H^\pm \to HW^*$ is possible if $m_{H^\pm} \gtrsim m_H$, and its decay rate is given by 
\begin{align}
\Gamma(H^\pm \to H W^{\pm *}) &\equiv \sum_f\Gamma(H^\pm \to H W^{\pm *}\to H f\bar{f}')=
\frac{9g^4}{512\pi^3}s_{\beta-\alpha}^2m_{H^\pm} G\left(z_{H},z_W^{*}\right). 
\end{align}

\begin{figure}[t]
 \begin{center}
\includegraphics[width=8cm]{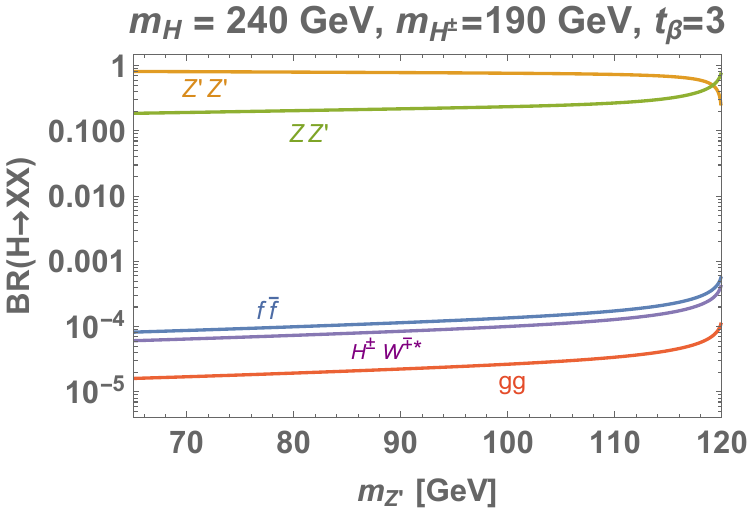}  \
\includegraphics[width=8cm]{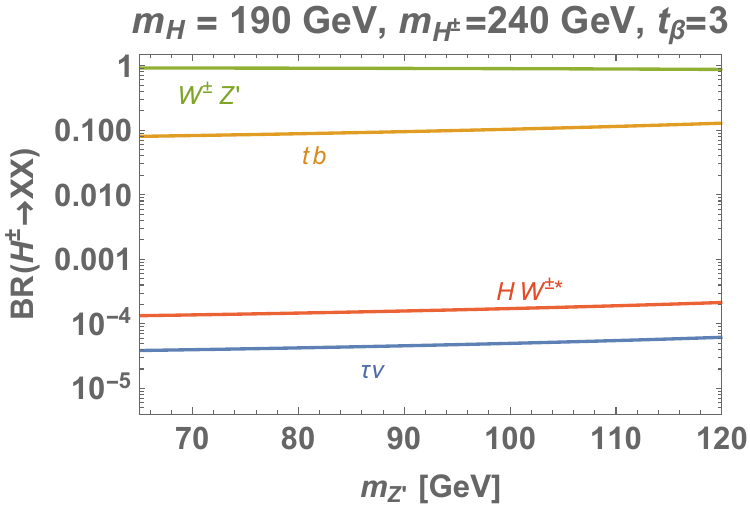}  
 \caption{Left: BRs of $H$ as a function of $m_{Z'}$ for $m_H = 240$ GeV, $m_{H^\pm}=190$ GeV and $\tan\beta = 3$. Right: BRs of $H^\pm$ as a function of $m_{Z'}$ for $m_H = 190$ GeV, $m_{H^\pm}=240$ GeV and $\tan\beta = 3$.   }
\label{fig:HBR}
\end{center}
\end{figure}

In Fig.~\ref{fig:HBR}, we show the BRs of $H$ (left panel) and $H^\pm$ (right panel) as a function of $m_{Z'}$ where $m_H = 240(190)$ GeV and $ m_{H^\pm} = 190(240)$ GeV in $H(H^\pm)$ decay, $t_\beta =3$ and $s_{\beta-\alpha} = 1$. 
We see that the $H \to Z'Z'/ZZ'$ modes are dominant, while the fermionic decay modes are negligibly small. 
Regarding the decay of the charged Higgs bosons, the $H^\pm \to W^\pm Z'$ and $H^\pm \to tb$ modes are about $90\%$ and $10\%$, respectively.

\begin{figure}[t]
 \begin{center}
\includegraphics[width=8cm]{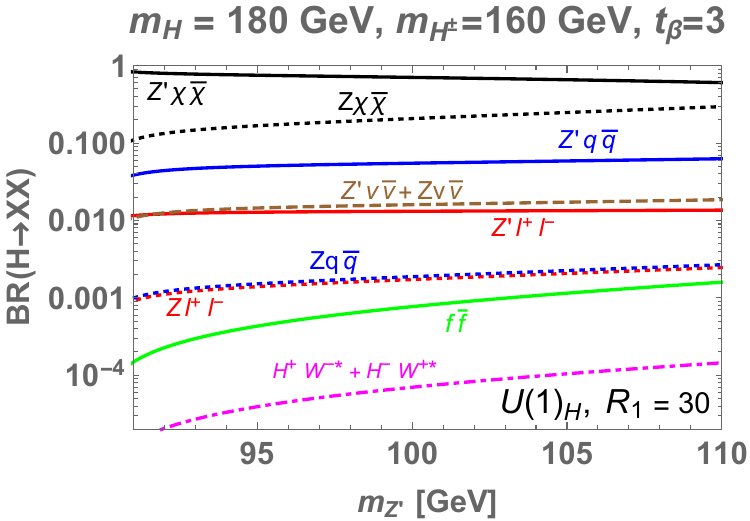}  
\includegraphics[width=8cm]{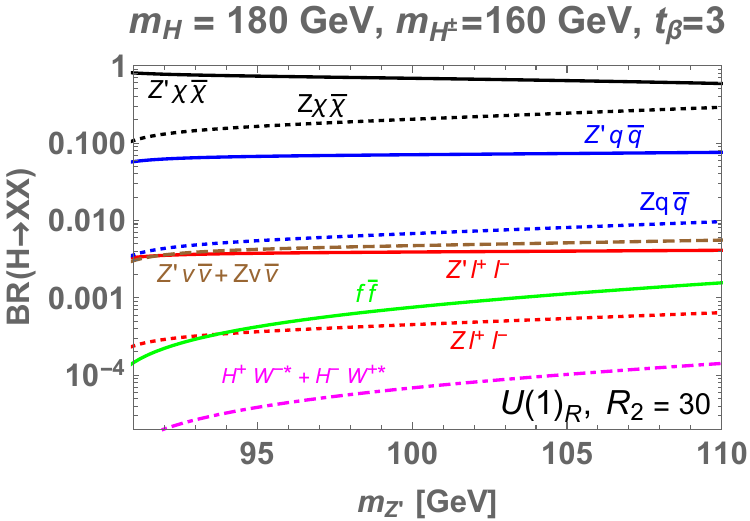}  
 \caption{Left and right panels respectively show the BRs of $H$ as a function of $m_{Z'}$ in the $U(1)_{H}$ model with $R_1=30$ and $U(1)_R$ model with $R_2 = 30$. 
 In both the plots, we take
 $m_H = 180$ GeV, $m_{H^\pm} = 160$ GeV and $\tan\beta =3$. }
\label{fig:HBR2}
\end{center}
\end{figure}

In Fig.~\ref{fig:HBR2}, we also show the BRs of lighter $H$ ($m_H =160$ GeV) as a function of $m_{Z'}$  for $t_\beta =3$ and $s_{\beta-\alpha} = 1$. 
In this case, three body decays of $H$ become dominant through $H$ decays into an off-shell and an on-shell gauge bosons, so that the final states depend on the $U(1)_X$ charges of fermions. 
The left and right panel show the case of $U(1)_H$ with $R_1=30$ and $U(1)_R$ with $R_2=30$, respectively. We see that the final states including $\chi$-pair are dominant in this case.

%%%%%%%%%%%%%%%%%%%%%%%%%%%%%%%%%%%%%%%%%

\section{Constraints \label{sec:const}} 

In this section, we discuss the constraints on the model parameters from theoretical bounds and experimental data. 

\subsection{Unitarity and vacuum stability}

We can apply similar constraints on the parameters of the Higgs potential
from perturbative unitarity~\cite{Kanemura:1993hm,Ginzburg:2005dt,Kanemura:2015ska} and vacuum stability~\cite{Deshpande:1977rw,Nie:1998yn,Kanemura:1999xf}, i.e., the condition for bounded from below as those given in the 2HDM with a $Z_2$ symmetry. 
Since the terms
\begin{align}
\frac{\lambda_5}{2}(\Phi_1^\dagger\Phi_2)^2 + \text{h.c.}\notag
\end{align}
are forbidden by the $U(1)_X$ symmetry, we can recast the expressions of these constraints without the $\lambda_5$ parameter.   
Therefore, the bounded from below condition is simply expressed as 
\begin{align}
\lambda_1>0,~~\lambda_2>0,~~
\sqrt{\lambda_1 \lambda_2}+\lambda_3+\operatorname{MIN}\left(0, \lambda_4\right)>0.
\end{align}
From Eq.~(\ref{eq:lam4}), the sign of $\lambda_4$ has to be negative. Thus, the last condition in the above can be rewritten as 
\begin{align}
    \sqrt{\lambda_1 \lambda_2}+\lambda_3-|\lambda_4|>0.
\end{align}
In the Higgs alignment limit, the above condition can be written by using Eqs.~(\ref{eq:lam1})-(\ref{eq:lam4}): 
\begin{align}
\sqrt{(m_h^2 + m_H^2t_\beta^2 )(m_h^2 + m_H^2/t_\beta^2)} +m_H^2-m_h^2 > 0.
\end{align}
Therefore, as long as $m_H^2 > 0$, the vacuum stability condition is satisfied.

For the perturbative unitarity bound, we impose 
\begin{align}
|\text{Re} (a_i)| \leq \frac{1}{2}, 
\end{align}
where $a_i$ are the eigenvalues of the $s$-wave amplitude matrix for $2\to 2$ elastic scatterings in the high-energy limit. 
In our model, $a_i$ are given by 
\begin{align}
& a_1^{\pm} = \frac{1}{32 \pi} \left[3(\lambda_1 + \lambda_2) \pm \sqrt{9 (\lambda_1 - \lambda_2)^2 + 4(2 \lambda_3 + \lambda_4)^2} \right], \\
& a_2^{\pm} = \frac{1}{32 \pi} \left[\lambda_1 + \lambda_2 \pm \sqrt{ (\lambda_1 - \lambda_2)^2 + 4\lambda_4^2} \right], \\
& a_3^{\pm} = \frac{1}{32 \pi} \left[\lambda_1 + \lambda_2 \pm \sqrt{(\lambda_1 - \lambda_2)^2 } \right], \\
& a_4 = \frac{1}{16 \pi} (\lambda_3 + 2 \lambda_4 ), \\
& a_{5} = \frac{1}{16 \pi} \lambda_3 , \\
& a_6 = \frac{1}{16 \pi} (\lambda_3 + \lambda_4).
\end{align}
These constraints can be converted into bounds on the Higgs boson masses and mixing angles via Eqs.~(\ref{eq:lam1})-(\ref{eq:lam4}). We find that $a_1^+$ provides the strongest constraint among the above eigenvalues. 
Thus, the maximally allowed value of $m_{H}^2$ is obtained by the following expression for fixed values of $m_{H^\pm}$ and $t_\beta$
\begin{align}
m_H^2(\text{max}) 
& = \frac{6M_0^2}{5}\Bigg[\xi^2-\frac{R}{6}
   - \sqrt{\left(\xi^2
 - \frac{R}{6} \right)^2
- \frac{5}{12^2}(R-1)(5-R) 
}\Bigg],
\label{eq:mH}
\end{align}
where 
\begin{align}
\xi &= \frac{t_\beta + 1/t_\beta}{2} ~(\geq 1),~~
R = \frac{4m_{H^\pm}^2 + 24\pi v^2 -5m_h^2}{M_0^2},~
    M_0^2 = 8\pi v^2 -3m_h^2\simeq 24v^2.
\end{align}
Since $\xi$ is symmetric under $t_\beta \leftrightarrow 1/t_\beta$, the bound for $0 <t_\beta < 1$ is obtained from that of $t_\beta > 1$ by the replacement of $t_\beta \to 1/t_\beta$. 
The $R$ value is constrained to be 
$3\lesssim R \leq 5$, where the lower bound is given by taking $m_{H^\pm}^2 = 0$. The upper bound on $R ~(=5)$ is obtained by noting that the second term inside the square root in Eq.~(\ref{eq:mH}) vanishes, by which $m_H^2\text{(max)}$ becomes zero.
This upper limit value $R = 5$ corresponds to $m_{H^\pm} \simeq 850$ GeV, which gives the absolute upper limit on $m_{H^\pm}$.  
We note that $m_H^2\text{(max)}$ takes its maximal value at $\xi = 1$, i.e., $t_\beta = 1$, for a fixed value of $m_{H^\pm}$ and it monotonically decreases as $\xi$ getting larger. 

\begin{figure}[tb]
 \begin{center}
\includegraphics[width=8cm]{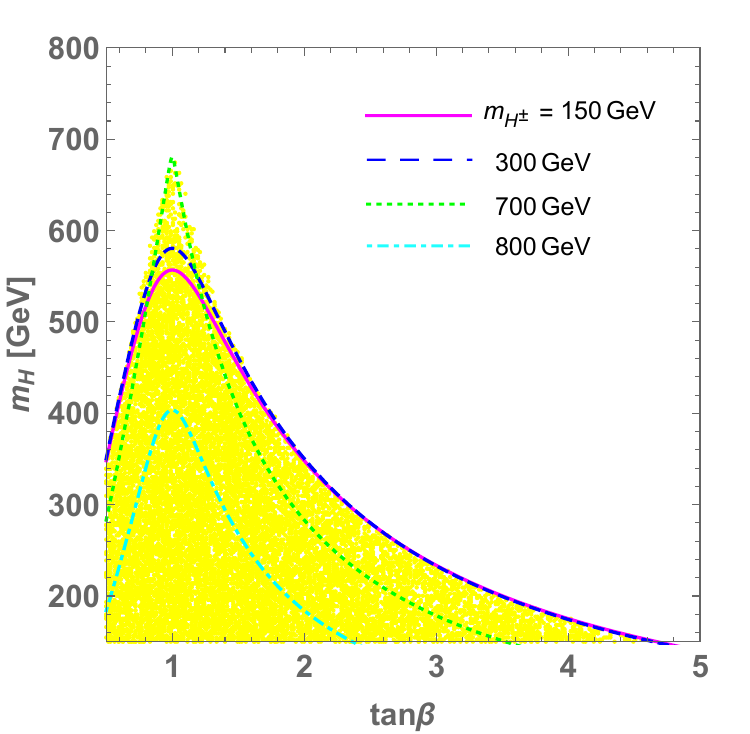}  \
\includegraphics[width=8.2cm]{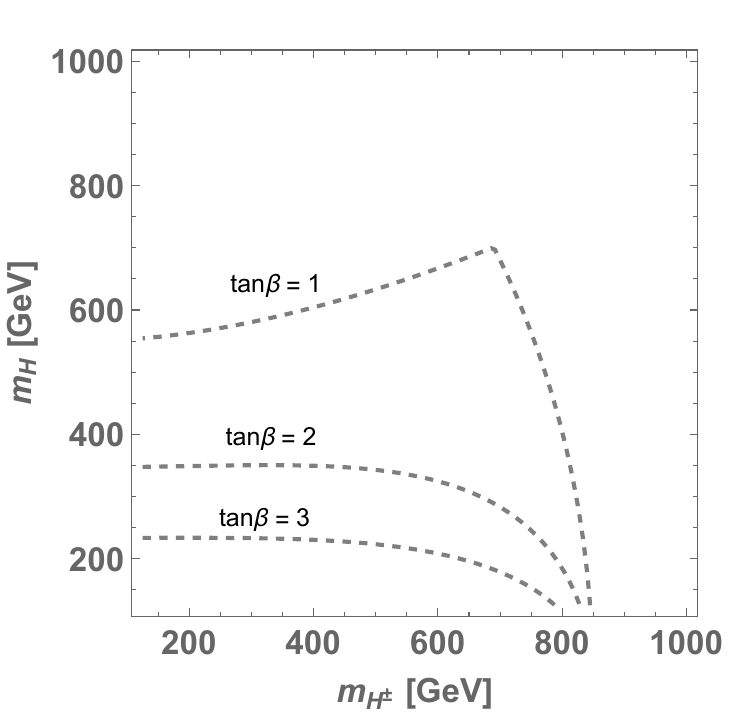}  
 \caption{Left: Yellow points show the allowed region on the $\{\tan \beta, m_H\}$ plane where magenta, blue-dashed, green-dotted, and cyan-dot-dashed curves correspond to the cases for  $m_{H^\pm}=150$ GeV, 300 GeV, 700 GeV and 800 GeV, respectively.
 Right: Upper bounds on $m_H$ as a function of $m_{H^\pm}$ for different values of $\tan \beta$ where region above each curve is excluded.
 }
\label{fig:unitarity}
\end{center}
\end{figure}

In the left panel of Fig.~\ref{fig:unitarity}, 
we show the upper limit on $m_H$ as a function of $t_\beta$ for 
$m_{H^\pm}=150$ GeV (solid, magenta), 300 GeV (dashed, blue), 700 GeV (dotted, green) and 800 GeV (dot-dashed, cyan) by using Eq.~\eqref{eq:mH}. 
The yellow shaded region is allowed by the perturbative unitarity bound which is obtained by scanning $m_{H,H^\pm} \in [150, 800]$ GeV. It is clear that the bound is most relaxed for $t_\beta = 1$ as we explained, and 
the maximally allowed value of $m_H$ is given at around 700 GeV. 
In the right panel, we also show the upper bound of $m_H$ as a function of $m_{H^\pm}$ for different values of $t_\beta$, where the region above each curve is excluded.

\subsection{Electroweak precision observables}

We impose bounds from the electroweak oblique parameters. Although we cannot simply apply the framework of the \( S \), \( T \) and \( U \) parameterization~\cite{Peskin:1990zt,Peskin:1991sw} to our model due to the existence of the $Z'$ boson, this formalism can be justified by assuming no $Z$-$Z'$ mixing at tree level and small coupling between $Z'$ and SM fermions. 
We then use the expressions of these oblique parameters given in the $Z_2$ symmetric version of the 2HDM~\cite{Branco:2011iw} as an approximation. Notice that the physical CP-odd Higgs boson is absent in our model due to the spontaneous breaking of the $U(1)_X$ symmetry, and it becomes the NG boson associated with $Z'$. Therefore, we obtain the values of the $S$, $T$ and $U$ parameters by replacing the mass of the CP-odd Higgs boson with $m_{Z'}$ in the ’t~Hooft-Feynman gauge.  
%The full expressions of the oblique parameters are summarized in the Appendix.~\ref{sec:STU}.
We use the values of the oblique parameters from the global fit of electroweak data~\cite{ParticleDataGroup:2024cfk} as follows
\begin{align}
S = -0.04 \pm 0.1,\quad 
T = 0.01\pm 0.12,\quad 
U = -0.01\pm 0.09. 
\end{align}

\subsection{Flavor constraints}

We take into account the following branching ratios of meson decays: 
\begin{align}
&{\cal B}(B_s \to X_s \gamma),\quad 
{\cal B}(B \to \tau\nu),\quad 
{\cal B}(B_s \to \mu^+\mu^-),\quad 
{\cal B}(B_d \to \mu^+\mu^-),\notag\\
&{\cal B}(D_s \to \mu\nu),\quad 
{\cal B}(D_s \to \tau\nu).
\end{align}
These processes are particularly sensitive to $m_{H^\pm}$ and $t_\beta$. 
Since our Yukawa interactions correspond to the Type-I structure, 
the $H^\pm$ contributions to these amplitude are suppressed by larger values of $m_{H^\pm}$ and/or $t_\beta$ due to the factor of $\cot^2\beta$. 
We employ the public code \texttt{SuperIso\_v4.1}~\cite{Mahmoudi:2008tp} to evaluate the flavor constraints at $2\sigma$ level in our model, which combines the experimental and theoretical uncertainties.  

\begin{figure}[tb]
\begin{center}
\includegraphics[width=14cm]{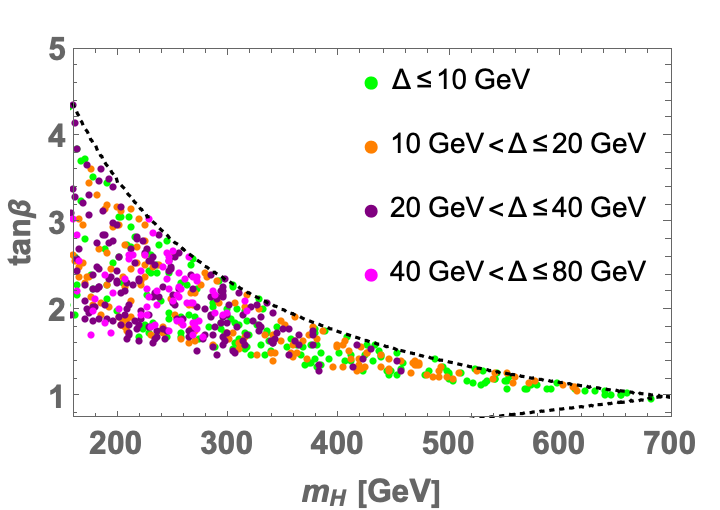}  \
\caption{Colored points are allowed by flavor constraints and electroweak precision data with different choice of the mass difference $\Delta \equiv |m_{H^\pm}-m_H|$. 
The dotted curves indicate upper and lower limits on $\tan\beta$ from the perturbative unitarity bound.
}
\label{fig:flavorDelta}
\end{center}
\end{figure}

In Fig.~\ref{fig:flavorDelta}, each point is allowed by the flavor constraints and the electroweak precision data, where the different color indicates the different values of the mass splitting $\Delta = |m_H - m_{H^\pm}|$, i.e., $\Delta\!\leq\!10~\mathrm{GeV}$ (green), $10~\mathrm{GeV}<\Delta\!\leq\!20~\mathrm{GeV}$ (orange), $20~\mathrm{GeV}<\Delta\!\leq\!40~\mathrm{GeV}$ (purple) and $40~\mathrm{GeV}<\Delta\!\leq\!80~\mathrm{GeV}$ (magenta).
We also show the upper and lower limits on $t_\beta$ from the perturbative unitarity bound by dotted curves. 
We see that the flavor constraint excludes smaller values of $t_\beta$, e.g., $t_\beta \lesssim 2$ (1)  for $m_H = 200$ (600) GeV.

\subsection{Constraints from $h$ decay}

As we have seen in Sec.~\ref{sec:decay}, $h$ can decay into $Z'Z^{\prime *}$ for $m_h/2 < m_{Z'}< m_h$. This induces 4 lepton final states as $h \to Z' \ell^+ \ell^- \to \ell'^+ \ell'^- \ell^+ \ell^-$ via a leptonic decay of $Z'$. 
Currently, such final states have been strongly constrained by the LHC data~\cite{ATLAS:2021wob}, by which the upper limit on the cross section has been taken to be $\sigma_{4 \ell} \simeq 0.1$ fb.
We thus require
\begin{equation}
\sigma(gg \to h) \times{\cal B}(h \to Z' \ell^+ \ell^-) \times {\cal B}(Z' \to \ell^+ \ell^-) <0.1 \ {\rm fb},\label{eq:4l}
\end{equation}
where $\sigma(gg \to h)$ indicates the Higgs production cross section by gluon fusion. 
We estimate the cross section by $\sigma(gg \to h)\simeq 50~\text{pb} \times \cot^2\beta$, where 50~\text{pb} is the cross section in the SM~\cite{Anastasiou:2016cez,Cepeda:2019klc}. 
From the above condition, 
the values of $m_{Z'}$ and $R_{1,2}$ are constrained. 

\begin{figure}[tb]
 \begin{center}
\includegraphics[width=8cm]{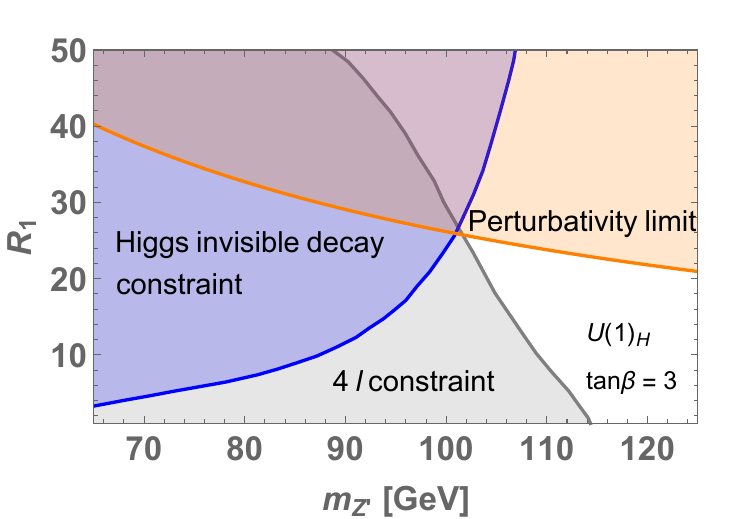}  \
\includegraphics[width=8cm]{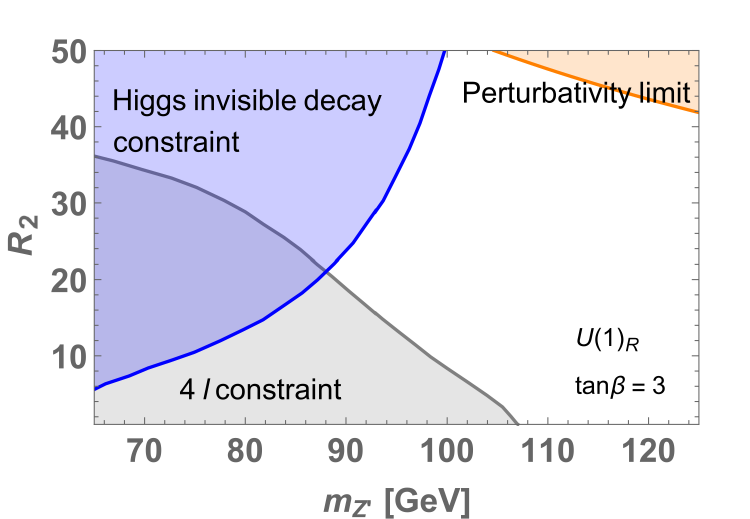}  
 \caption{Left plot shows the region excluded on the $\{m_{Z'},R_1\}$ plane by 4 lepton searches (gray shaded), pertabativity of the $Z' \chi \bar{\chi}$ coupling (orange shaded) and the constraint from Higgs invisible decay (blue shaded) in the $U(1)_H$ model with 
  $\tan \beta =3$. 
 The right plot is similar to the left one  but shows the constraints on the $\{m_{Z'},R_2\}$ plane in the $U(1)_R$ model.
 }
\label{fig:4lconst-h}
\end{center}
\end{figure}

In Fig.~\ref{fig:4lconst-h}, 
the shaded region is excluded by the constraint from Eq.~(\ref{eq:4l}) on the 
$m_{Z'}$-$R_1$ plane in the $U(1)_H$ model (left panel) and on the 
$m_{Z'}$-$R_2$ plane in the $U(1)_R$ model (right panel).
Here, we also take into account the perturbativity limit for the $Z' \chi \bar{\chi}$ coupling requiring $R_{1,2} g_{\chi \chi Z'} < \sqrt{4 \pi}$. The orange-dashed curves in Fig.~\ref{fig:4lconst-h} indicate the upper bound of $R_{1,2}$ for different $t_\beta$ values.
In addition, we show excluded region by Higgs invisible decays via $h \to Z' \chi \bar{\chi} \to (\chi \bar{\chi})(\chi \bar{\chi})$, adopting ${\cal B}(h \to {\rm invisible}) < 0.107$~\cite{ATLAS:2023tkt}, by blue shaded region.
We see that the wider region is allowed in the $U(1)_R$ model than that in the $U(1)_H$ model, because the leptonic decay mode of $Z'$ is slightly suppressed by the choice of the $U(1)_X$  charges. 

Apart from the $h$ decays into new modes, 
the $h\to \gamma\gamma$ decay mode can significantly be modified from the SM prediction due to the non-decoupling contribution from the charged Higgs boson loop~\cite{Mondal:2025iie}.   
The decay rate is expressed as
\begin{align}
&\Gamma(h \to \gamma\gamma)=  \frac{\sqrt{2}G_F\alpha_{\text{em}}^2m_h^3}{16\pi^3 }  \left|(t,W)_{\rm SM} + \frac{v\lambda_{H^+H^-h}}{2m_h^2}\left[1 - \frac{4m_{H^\pm}^2}{m_h^2} \arcsin^2\left(\frac{m_h}{2m_{H^\pm}}\right)\right]
\right|^2,
\end{align}
where $(t,W)_{\rm SM}$ indicates the top and W boson loop contributions in the SM, and 
its numerical value is about $1.6$. The $H^+H^-h$ coupling $\lambda_{H^+H^-h}$ is given 
in the Higgs alignment limit and for $m_{H^\pm} > m_h/2$:  
\begin{align}
   \lambda_{H^+H^-h} = \frac{1}{v}(2m_{H^\pm}^2 + m_h^2). \label{eq:lamhphmh} 
\end{align}
For $m_h^2/m_{H^\pm}^2 \ll 1 $, the decay rate is approximately given by 
\begin{align}
&\Gamma(h \to \gamma\gamma)\simeq  \frac{\sqrt{2}G_F\alpha_{\text{em}}^2m_h^3}{16\pi^3 }  \left|(t,W)_{\rm SM} -\frac{1}{12} - \frac{19}{360}\frac{m_h^2}{m_{H^\pm}^2} 
\right|^2. 
\end{align}
We see that the effect of the charged Higgs loop does not vanish but it becomes constant and destructive with respect to the SM part $(t,W)_{\rm SM}$. 
By comparing the current data for the signal strength value $\mu_{\gamma\gamma} = 1.04^{+0.10}_{-0.09}$~\cite{ATLAS:2022tnm}, we obtain the lower limit on $m_{H^\pm}$ to be about 160 GeV at 95\% CL. 

\subsection{Constraints from the search for 4 lepton channels}

\begin{figure}[tb]
 \begin{center}
\includegraphics[width=8cm]{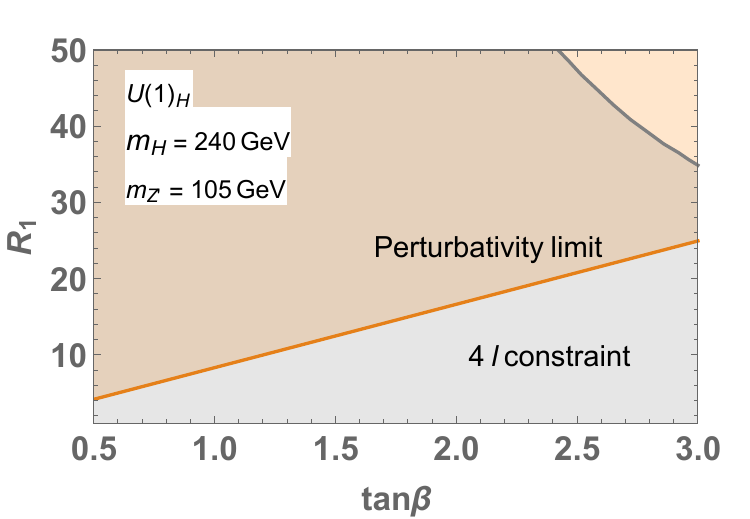}  \
\includegraphics[width=8cm]{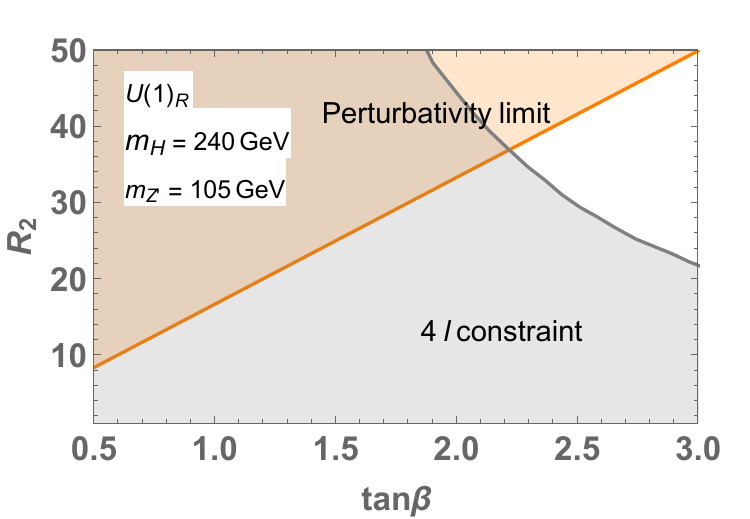}  
\includegraphics[width=8cm]{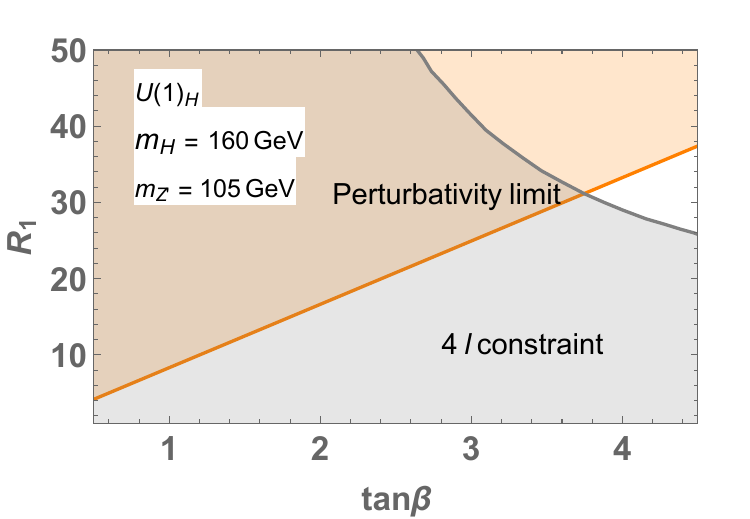}  \
\includegraphics[width=8cm]{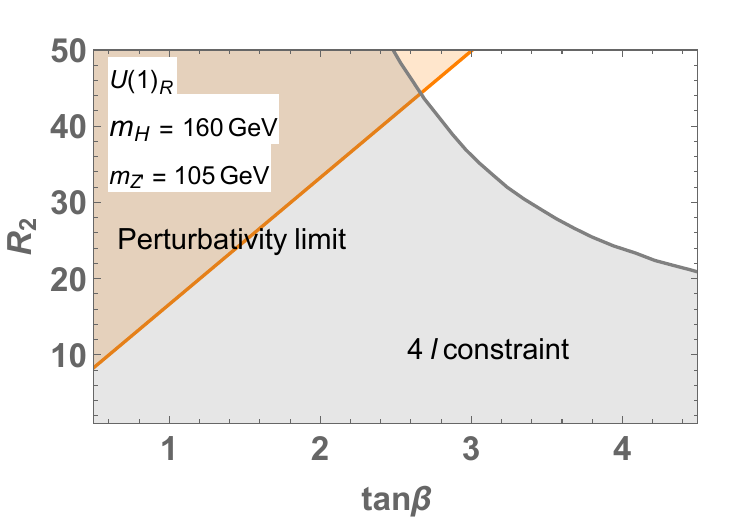}  
 \caption{Left plots show the region excluded on the $\{\tan\beta,R_1\}$ plane by 4 lepton searches (gray shaded) and the pertabativity of the $Z' \chi \bar{\chi}$ coupling (orange shaded) in the $U(1)_H$ model with $m_H=240$ GeV (upper panel) and $m_H = 160$ GeV (lower panel). 
 The right plots are similar to the left ones but show the constraints on the $\{\tan\beta,R_2\}$ plane in the $U(1)_R$ model.
 We take $m_{Z'} = 105$ GeV for all the plots. 
   }
\label{fig:4lconst-Heavy}
\end{center}
\end{figure}

As we have seen in Sec.~\ref{sec:decay}, $H$ dominantly decays into $Z'Z'$ and $Z'Z$ leading to 4 lepton final states from the leptonic decays of $Z'$ and $Z$. 
We impose the upper limit on the cross section based on~\cite{ATLAS:2021wob}
\begin{align}
& \sigma(gg \to H) \times {\cal  B}(H \to 4\ell) < 0.1 \ {\rm fb}, 
\end{align}
where $\sigma(gg\to H)$ indicates the cross section of $H$ via the gluon fusion process, and ${\cal B}(H \to 4\ell)$ denotes the branching ratio of $H$ into four lepton final states which are induced via $Z^{\prime(*)} \to \ell^+\ell^-$ and $Z^{(*)} \to \ell^+\ell^-$.

In Fig.~\ref{fig:4lconst-Heavy}, we show the constraints on the parameter space from the 4 lepton search and the perturbatively on the $t_\beta$-$R_1$ and $t_\beta$-$R_2$ plane in the $U(1)_H$ model and the $U(1)_R$ model, respectively. 
We take $m_{Z'} = 105$ GeV and $m_H = 240$ GeV (upper panels) and 160 GeV (lower panels). 
It is shown that $R_{1,2}$ should be $\mathcal{O}(20)$ or larger to avoid the constraint from four lepton search.
In addition, the $U(1)_H$ model is more severely constrained as compared with the $U(1)_R$ model due to the larger branching ratio of $Z' \to \ell^+\ell^-$. 

\subsection{Searches for $Z'$}

We briefly summarize the relevant constraints from searches for $Z'$. 
For details, see e.g.,~\cite{Ilten:2018crw, Bauer:2018onh, Asai:2022zxw, Asai:2023mzl, KA:2023dyz,Chakraborty:2021apc}.
\begin{itemize}
  \item LEP limit: The existence of $Z'$ modifies the cross section of $e^+ e^- \to f \bar{f}$ process due to the $s$-channel mediation of $Z'$.
We adopt the $e^+ e^- \to \mu^+ \mu^-$ cross section data at the $Z$ pole as the most stringent constraint that is $\sigma_{\rm exp}(e^+e^-\to \mu^+ \mu^-) = 2.0018 \pm 0.0060$ nb with $\sqrt{s} =91.23$ GeV~\cite{Electroweak:2003ram, ALEPH:1997gvm}. 
We then require that the cross section in our model $\sigma(e^+e^- \to \mu^+ \mu^-)$ is within $2 \sigma$ error of $\sigma_{\rm exp}(e^+e^-\to\mu^+ \mu^-)$, by which the kinetic mixing parameter $\epsilon$ (gauge coupling $x_R g_X$) and the $Z'$ mass are constrained in the $U(1)_H$ ($U(1)_R$) model. 
\item Dark photon searches at LHC: The LHCb~\cite{LHCb:2019vmc} and CMS~\cite{CMS:2019kiy,CMS:2023slr} experiments are searching for a dark photon $A'$ which decays into a muon pair $\mu^+ \mu^-$. The data from these experiments constrain the kinetic mixing parameter $\epsilon_{\rm DP}$ and dark photon mass $m_{A'}$ in the dark photon case. 
We estimate an upper limit of our kinetic mixing parameter $\epsilon^{\rm max}$ 
and the coupling $[x_R g_X]^{\rm max}$ in the $U(1)_H$ and $U(1)_R$ cases, respectively, by re-scaling the upper limit of $\epsilon_{\rm DP}$ as follows:
\begin{equation}
X^{\rm max} = \epsilon^{\rm max}_{\rm DP}(m_{A'}) \sqrt{\frac{\tilde \sigma(pp \to A') {\cal B}(A' \to \mu^+ \mu^-)}{\tilde \sigma(pp \to Z')  {\cal B}(Z' \to \mu^+ \mu^-)}},
\end{equation}
where 
$X^{\rm max}$= $\epsilon^{\rm max}$ and $[x_R g_R]^{\rm max}$ in the $U(1)_H$ and $U(1)_R$ cases, respectively. In the above equation, 
$\tilde \sigma(pp \to A'/Z')$ is the $A'/Z'$ production cross section with $\epsilon_{\rm DP}$, $\epsilon$, and $x_R g_R$ being taken to be unity, and $\epsilon^{\rm max}_{\rm DP}(m_{A'})$ is the upper bound of the kinetic mixing parameter from the dark photon search as a function of $m_{A'}$.
The cross sections are estimated by {\tt CalcHEP~3.8}~\cite{Belyaev:2012qa} implementing relevant interactions.

\item Neutrino scattering experiments: 
$e$-$\nu$ scatterings can constrain $Z'$ mass and $\epsilon$ in the $U(1)_H$ model since $Z'$ interacts with both electrons and neutrinos. We can estimate the upper limit of $t_\epsilon$ from the experimental data by calculating the $e$-$\nu$ scattering cross section~\cite{Asai:2023mzl,KA:2023dyz}. In our model, it is found that constraints from $e$-$\nu$ scattering is weaker than the LEP limit and we do not discuss details in this paper. Note also that $Z'$ does not contribute to the $e$-$\nu$ scattering in the $U(1)_R$ model. 
\end{itemize}

\begin{figure}[tb]
 \begin{center}
\includegraphics[width=8cm]{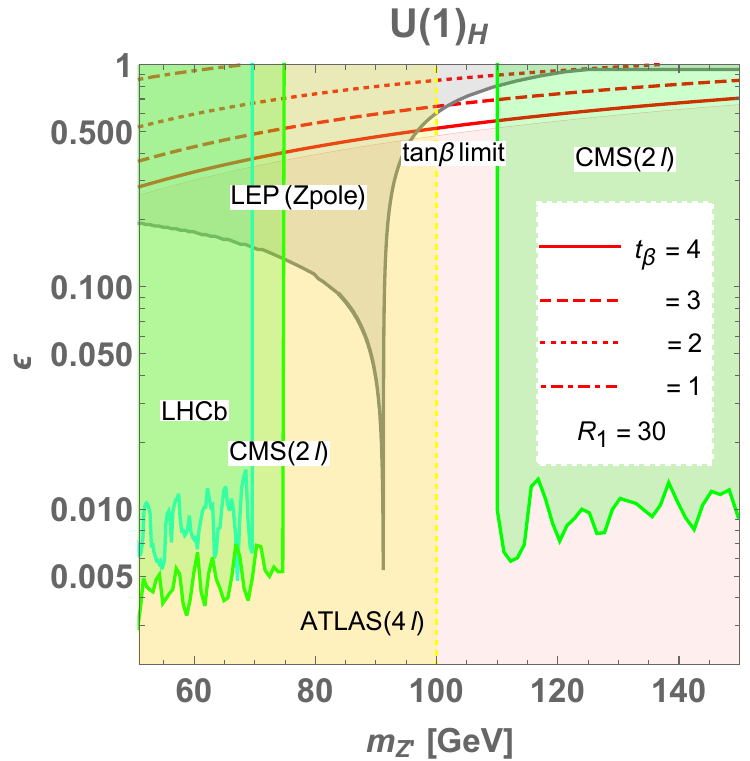}  \
\includegraphics[width=8cm]{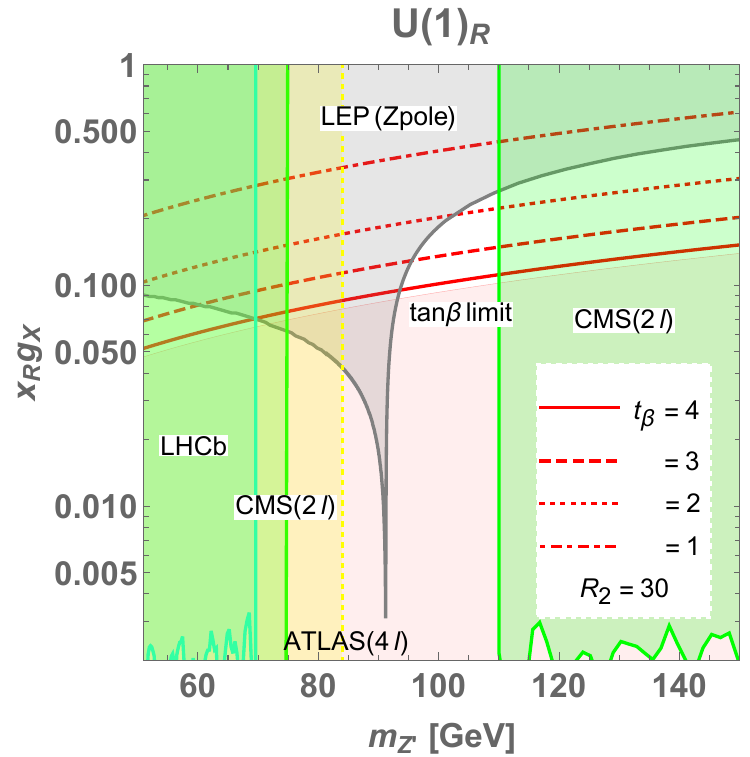}  
 \caption{Left plot shows the region excluded by constraints from the $Z'$ searches and the constraint from $h \to 4\ell$ on the $\{m_{Z'},\epsilon\}$ plane in the $U(1)_H$ model with $R_1 = 30$. 
 Each red curve shows the value of $\tan\beta$ and the region below the light-red curve (indicated by ``$\tan\beta$ limit") is excluded by the perturbative unitarity bound. 
 The right plot is similar to the left one but the constraints on the $\{m_{Z'},x_R^{}g_X^{}\}$ plane in the $U(1)_R$ model with $R_2 = 30$. }
 \label{fig:ZpConst}
\end{center}
\end{figure}

The left and right panels in Fig.~\ref{fig:ZpConst} show the summary of excluded regions on the $\{m_{Z'}, \epsilon\}$ and $\{m_{Z'},x_R g_X\}$ plane in the $U(1)_{H}$ model with $R_1=30$ and the $U(1)_R$ model with $R_2 = 30$, respectively. 
The color shaded regions are excluded by the dark photon searches at LHCb (cyan), those at CMB (green), 
the LEP constraint (gray) and the $h \to 4 \ell$ at LHC (yellow). 
The red curves indicate relation between $m_{Z'}$ and $\epsilon~(x_R g_X)$ under the no $Z$-$Z'$ mixing condition given in Eqs.~\eqref{eq:no-mixing-U1H} and \eqref{eq:no-mixing-U1R}. The light-pink region is excluded due to our upper bound on $\tan \beta$ from perturbative unitarity constraint. 
Thus, we find allowed region around $m_{Z'} \in [96, 110]$ GeV and $\epsilon \in [0.46, 0.78]$ in the $U(1)_H$ model and $m_{Z'} \in [93, 110]$ GeV and $\epsilon(x_R g_X) \in [0.087, 0.25]$ in the  $U(1)_{R}$ model. 
These regions can be further explored in future collider experiments such as Z-factories, e.g. FCC-ee~\cite{FCC:2018byv} and CEPC~\cite{CEPC-SPPCStudyGroup:2015csa,CEPCStudyGroup:2018ghi,CEPCStudyGroup:2023quu,Ai:2025cpj,CEPCStudyGroup:2025kmw}, providing more precision test around Z-pole.

\begin{figure}[tb]
 \begin{center}
\includegraphics[width=8cm]{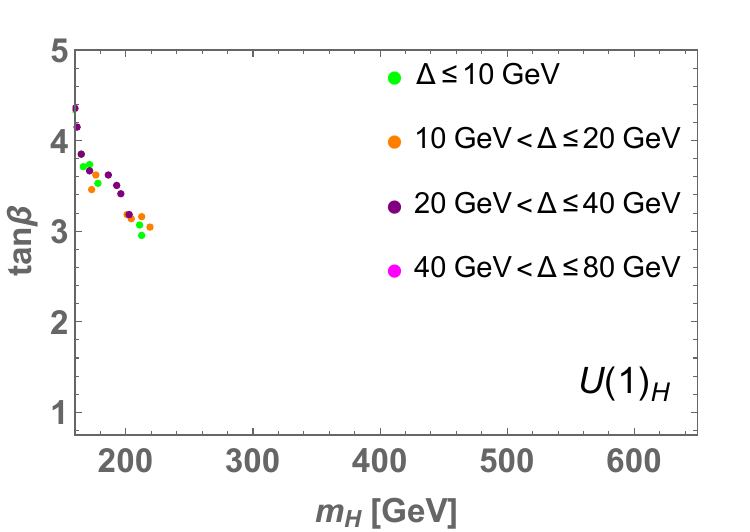}  \
\includegraphics[width=8cm]{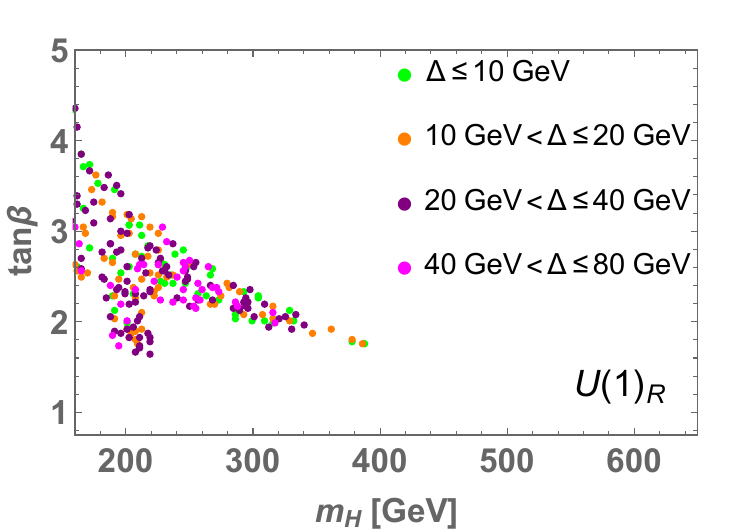}  
 \caption{Left and right panels respectively show allowed points on the $\{m_H, \tan \beta\}$ plane adopting all the constraints discussed in Sec.~\ref{sec:const} in the $U(1)_{H}$ and $U(1)_R$ models. 
 The meaning of colors is the same as those in Fig.~\ref{fig:flavorDelta}. 
  }
\label{fig:ConstComb}
\end{center}
\end{figure}

\section{Combined constraints \label{sec:combined}}

In this section, we combine all the constraints discussed in the previous section and show the allowed parameter region.

The left (right) panel in Fig.~\ref{fig:ConstComb} shows the allowed points on the $\{m_H, t_\beta\}$ plane adopting the constraints from vacuum stability, perturbative unitarity, oblique parameters, flavor data, and $h/H\to 4\ell$ in the $U(1)_H$ ($U(1)_R$) case, where the meaning of colors is the same as those in Fig.~\ref{fig:flavorDelta}.
Note that the parameters $m_{Z'}$ and $R_{1,2}$ are scanned over 
\begin{equation}
m_{Z'} \in [90, 110] \ {\rm GeV}, \quad R_{1,2} \in [1, R_{1,2}^{\rm max}],
\end{equation}
where $R_{1,2}^{\rm max}$ are determined by the perturbativity limit for the $Z' \chi \bar{\chi}$ coupling as 
\begin{align}
R_1^{\rm max} &=\sqrt{4\pi} \frac{v t_\beta }{m_{Z'}},\quad
R_2^{\rm max} =\sqrt{4\pi}\frac{2 v t_\beta }{m_{Z'}}, 
\end{align}
which are derived from Eqs.~(\ref{eq:no-mixing-U1H}), (\ref{eq:no-mixing-U1R}) and (\ref{eq:zpchichi}). 
We find that the quite limited region with $m_H \lesssim 220$ GeV and $3 \lesssim t_\beta \lesssim 4$
is allowed in the $U(1)_H$ model, where the region with $m_H \gtrsim 220$ GeV and $\tan \beta \lesssim 3$ is excluded by the constraints from 4 lepton decay of $H$.
On the other hand, relatively wider region of the parameter space is allowed in the $U(1)_R$ model compared with the $U(1)_H$ model, where the region with $m_H \gtrsim 400$ GeV and $t_\beta \lesssim 3$  is excluded by the 4 lepton decay of $H$.
We note that $t_\beta \lesssim 2.5$ and $m_H \lesssim 190$ GeV region is excluded by the perturbative limit since we need larger $R_{1,2}$ to suppress four lepton production cross section against the increase of $H$ production cross section in both the models. 

These limited regions of the parameter space can further be explored at future collider experiments such as the High-Luminosity LHC and lepton colliders, e.g., ILC~\cite{Baer:2013cma,Asai:2017pwp,Fujii:2017vwa,ILC:2019gyn}, CEPC~\cite{CEPC-SPPCStudyGroup:2015csa} and FCC-ee~\cite{FCC:2018byv}. 

\section{Conclusions \label{sec:conclusions}}

We have discussed 2 Higgs doublet models (2HDMs) with a new $U(1)$ gauge symmetry, denoted as $U(1)_X$, which induces a new neutral gauge boson $Z'$. In those models, we also introduced a pair of vector-like fermions $\chi$ which are SM-singlet but charged under $U(1)_X$ in order to avoid constraints from Higgs boson decays into four leptons at LHC by modifying the branching ratio of the $Z'$ boson. 
In addition, $\chi$ could be a dark matter candidate whose stability is guaranteed by the $U(1)_X$ symmetry. 
Furthermore, we imposed the condition to cancel the $Z$-$Z'$ mixing to avoid constraints from precision measurement of the $Z$ boson interactions. 
As concrete benchmark models, we have considered the $U(1)_H$ and $U(1)_R$ models, where $Z'$ interactions with SM particles are induced only via the kinetic-mixing in the former while 
those with left-handed fermions vanish due to the $U(1)$ charge assignment in the latter.  

The Higgs sector of our scenario corresponds to more constrained version of a usual $Z_2$ symmetric case with a soft-breaking term, i.e., no $\lambda_5$ term in the potential. 
In addition, a physical CP-odd Higgs boson does not appear because it would be a longitudinal component of $Z'$. The Yukawa interactions take the same form as those given in the Type-I 2HDM. 
We have imposed the Higgs alignment limit, by which the couplings for the SM-like Higgs boson $h$ become the same as those of the SM predictions at tree level. 

One of the most important differences from models with $U(1)_X$ which is spontaneously broken by the vacuum expectation value (VEV) of a SM-singlet scalar field is the appearance of the upper limit on the masses of $Z'$ and extra Higgs bosons ($H$ and $H^\pm$) due to the fact that both $U(1)_X$ and the  electroweak symmetry are broken by the electroweak VEVs. Thus, their masses are typically constrained to be smaller than the TeV scale.   
Another significant difference can be seen in the $h$ decays, i.e., $h \to Z'Z'$ can easily dominate the branching ratio of $h$ in our models, and it cannot be suppressed by taking the no $Z$-$Z'$ mixing and the Higgs alignment limit. 
Such a large decay width of $h \to Z'Z'$ can only be avoided by taking a larger mass of $Z'$ ($m_{Z'} > m_h/2$) such that the on-shell decay of $h \to Z'Z'$ is kinematically forbidden. 

We then have taken into account various constraints on the parameter space under the no $Z$-$Z'$ mixing condition and the Higgs alignment limit.
As for theoretical constraints, we have imposed the vacuum stability, perturbative unitarity 
and perturbativity of the $Z'\bar{\chi}\chi$ interaction. 
We found that the perturbative unitarity constraint sets a severe upper limit on the masses of $H$ and $H^\pm$ which are analytically expressed in Eq.~(\ref{eq:mH}). We obtained $m_H \lesssim 700$ GeV and 
$m_{H^\pm} \lesssim 850$ GeV. 
We have also taken into account the constraints from oblique parameters, flavor data, signal significance of $h \to \gamma\gamma$ and $h/H \to 4\ell$ as well as the various $Z'$ searches. 
Among the above, the decay width of $h \to \gamma \gamma$ is significantly modified due to the $H^\pm$ loop contribution, and the current LHC data require the mass of $H^\pm$ to be heavier than around 160 GeV.
In addition, the cross section of the $h/H \to 4\ell$ processes are enhanced due to the new decay channels of $h/H \to Z'Z'$ and $H \to ZZ'$ where one of the gauge bosons can be an off-shell mode.   
These constraints from 4 lepton final states can be avoided by considering the scenario where
$Z'$ dominantly decays into $\chi \bar{\chi}$ (missing final states) which can be realized by taking  larger values of $\chi \bar{\chi}Z'$ interaction under the perturbativity limit. 
Without introducing $\chi$, the models are completely excluded by the constraint.

Finally, we have combined all the constraints discussed above. 
We have found that the allowed region is given to be
$m_{Z'} \in [96, 110]$ GeV and $\epsilon \in [0.46, 0.78]$ in the $U(1)_H$ model with $R_1 = 30$ and 
$m_{Z'} \in [93, 110]$ GeV and $x_R g_X \in [0.087, 0.25]$ in the $U(1)_{R}$ model with $R_2 = 30$ as summarized in Fig.~\ref{fig:ZpConst}.
Furthermore, the allowed region has been found to be $m_H \in [160, 220]$ GeV and $\tan \beta \in [3, 4.4]$ in the $U(1)_H$ model and 
$m_H \in [160, 380]$ GeV and $\tan \beta \in [1.6, 4.4]$ for $U(1)_R$ model as summarized in Fig.~\ref{fig:ConstComb}.

Therefore, the allowed region in the models is strongly limitted by current phenomenological data, and the full region could be tested in future experiments such as the HL-LHC, the ILC, the FCC-ee, and the CEPC experiments. 

\vspace*{-4mm}

%%%%%%%%%%%%%%%%%%%%%%%%%%%%%%%%%%%
\section*{Acknowledgments}
%\vspace{0.3cm}
The work was supported by the Fundamental Research Funds for the Central Universities (T.~N.). Part of this work was completed before the authors (Y.~L. and X.~X.) left Sichuan University.
%%%%%%%%%%%%%%%%%%%%%%%%%%%%%%%%%%%

\bibliography{references}

\end{document}